\documentclass[superscriptaddress,aps,prb,twocolumn,amsmath,amssymb]{revtex4-1}

\usepackage{graphicx}
\usepackage{dcolumn}
\usepackage{color,soul}
\usepackage{multirow}
\usepackage{longtable}
\usepackage{float}
\usepackage{textcomp}
\usepackage[colorlinks=true, citecolor=red, urlcolor=blue]{hyperref}

\begin{document}

\title{The conundrum of relaxation volumes in first-principles calculations of charge defects}

\author{Anuj Goyal}
\email{agoyal@mines.edu}
\altaffiliation{Current Address: Metallurgical and Materials Engineering, Colorado School of Mines, Golden, CO 80401, USA}
\affiliation{Department of Materials Science and Engineering, University of Florida, Gainesville, FL 32611, USA}

\author{Kiran Mathew}
\affiliation{Department of Materials Science and Engineering, University of Florida, Gainesville, FL 32611, USA}

\author{Richard G. Hennig}
\affiliation{Department of Materials Science and Engineering, University of Florida, Gainesville, FL 32611, USA}

\author{Aleksandr Chernatynskiy}
\affiliation{Department of Physics, Missouri University of Science and Technology, Rolla, MO 65409, USA}

\author{Christopher R. Stanek}
\affiliation{Materials Science and Technology Division, Los Alamos National Laboratory, Los Alamos, NM 87545, USA}
  
\author{Samuel T. Murphy}
\affiliation{Engineering Department, Lancaster University, Brailrigg, Lancaster, LA1 4YW, UK}
  
\author{David A. Andersson}
\affiliation{Materials Science and Technology Division, Los Alamos National Laboratory, Los Alamos, NM 87545, USA}

\author{Simon R. Phillpot}
\affiliation{Department of Materials Science and Engineering, University of Florida, Gainesville, FL 32611, USA}

\author{Blas P. Uberuaga}
\email{blas@lanl.gov}
\affiliation{Materials Science and Technology Division, Los Alamos National Laboratory, Los Alamos, NM 87545, USA}

\date{\today}

\begin{abstract}
The defect relaxation volumes obtained from density-functional theory (DFT) calculations of charged vacancies and interstitials are much larger than their neutral counterparts, seemingly unphysically large. In this work, we investigate the possible reasons for this and revisit the methods that address the calculation of charged defect structures in periodic DFT. We probe the dependence of the proposed energy corrections to charged defect formation energies on relaxation volumes and find that corrections such as the image charge correction and the alignment correction, which can lead to sizable changes in defect formation energies, have an almost negligible effect on the charged defect relaxation volume. We also investigate the volume for the net neutral defect reactions comprised of individual charged defects, and find that the aggregate formation volumes have reasonable magnitudes. This work highlights an important issue that, as for defect formation energies, the defect formation volumes depend on the choice of reservoir. We show that considering the change in volume of the electron reservoir in the formation reaction of the charged defects, analogous to how volumes of atoms are accounted for in defect formation volumes, can renormalize the formation volumes of charged defects such that they are comparable to neutral defects. This approach enables the description of the elastic properties of isolated charged defects within the overall neutral material, beyond the context of the overall defect reactions that produce the charged defect.
\end{abstract}



\maketitle

\section{\label{sec:I}Introduction}

Charged defects in nonmetals, such as semiconductors and insulators, play a central role in determining electronic and electrical behavior of materials for semiconductor and battery applications\cite{Seebauer2006}. Charged defects are also very relevant to materials under extreme conditions of temperature, pressure, and irradiation, as they play a major role in accommodating stoichiometry variations\cite{Was2007}. In addition, defects induce lattice distortions in the perfect crystal, leading to finite changes in volume which can, at least in principle, be experimentally measured and compared to theoretical calculations\cite{Gillan1981, Lidiard1981, Nazarov2016}. These lattice distortions are particularly important in determining the interaction of defects with external strains, dislocations, and grain boundaries\cite{Clouet2008, Goyal2015}, and play a central role in the evolution of microstructures in materials.\par

The defect formation volume is an intrinsic property of the defect; theory for computing the volume change due to point defects gained attention in the early 1980\textquotesingle s with a focus on point defects in ionic crystals\cite{Gillan1981, Lidiard1981}. It is important to note that experimental measurements of the volume change associated with point defects yield the defect-averaged response of a mass and charge balanced system\cite{Gillan1981, Lidiard1981}; as a result there is little understanding of the volume change or lattice distortion associated with individual charged defects. Surprisingly, compared to the vast literature on the formation energy of charged defects, there have been relatively few theoretical investigations of their formation and relaxation volumes\cite{}. Investigations of the formation volume of charged point defects in Si\cite{Centoni2005, Windl2002a, Bruneval2012, Bruneval2015}, ZnO\cite{Erhart2006}, In$_2$O$_3$\cite{Agoston2009}, CeO$_2$\cite{Grieshammer2013} indicate that the defect formation volume is a strong function of the charge state of the defect, and as electrons are added (or removed) to the system the formation volume increases (or decreases) linearly with charge.\par

Defect formation and relaxation volumes are closely related concepts. The defect relaxation volume is simply the change in volume of a computational cell due to relaxation of the crystal lattice around the defect site. By contrast, the defect formation volume is calculated as the change in the volume associated with the formation reaction for the defect; it, therefore, accounts for the reference volume associated with reservoir to or from which atom(s) are moved in order to create the defect. We describe defect relaxation and formation volume and the differences between the two in more detail in Sec. \ref{sec:IIB}.\par

The theory of relaxation volume and lattice relaxation around the defect site can be used to obtain the so-called defect elastic dipole tensor\cite{Gillan1984, Nowick1972}, which relates the atomic structure of the point defect with its elastic distortion of the lattice. The elastic dipole tensor\cite{Gillan1984, Freedman2009, Nazarov2016}, $G_{ij}$, captures the symmetry of the point defect and is given as

\begin{equation}\label{eq:1}
G_{ij} = C_{ijkl}\Delta\Omega_{kl}^{r}
\end{equation}
where $C_{ijkl}$ is the elastic constant tensor, and $\Omega_{kl}^{r}$ is the relaxation volume tensor of the defect\cite{Puchala2008}. The elastic dipole tensor describes the change in formation energy of the defect, to first order, in the presence of external strains\cite{Clouet2008}. Therefore, computation of the defect relaxation volume tensor quantifies the lattice distortion around the defect and thus the elastic interaction with other defects and strain fields\cite{Hinterberg2013, Goyal2015}.\par

Building on our previous work on point defect behavior in the presence of external strain\cite{Goyal2015}, here we find that the relaxation volumes of individual charged point defects in uranium dioxide (UO$_{2}$) to be much larger than those of the corresponding neutral defects. Further, when we calculate the corresponding elastic dipole tensor, the predicted changes in energy of the defect under elastic strain fields are extremely large; indeed, they are so large that they seem to be unphysical. As the dipole tensor is directly related to defect relaxation volumes, this result motivates us to examine this issue in greater depth.\par

This work addresses the issue of the relaxation volume of charged defects, using UO$_{2}$ as a prototypical material, both because of its complex electronic structure and because its fluorite crystal structure is common to many technologically important materials. Defect calculations\cite{Dorado2011, Andersson2011, Dorado2012, Nerikar2009} in UO$_{2}$ have established that defects such as Frenkel and Schottky defect pairs and fission product complexes are likely to consist of individual components that are charged. The individual volume change depends not only on the size mismatch of the defect species, but also on the more complex charge transfer mechanisms occurring in the presence of the defect. In addition to UO$_{2}$, we present results and comparisons for relaxation volumes of charged defects in Si and GaAs, for the purpose of validating against existing data in these systems and uncovering behavior that applies to charged systems more generally.\par

The paper is arranged as follows. Section \ref{sec:IIA} presents the density-functional theory (DFT) methodology used in calculating defect properties. In Secs. \ref{sec:IIB} and \ref{sec:IIC}, we discuss the definitions of defect formation and relaxation volumes. Various schemes to correct the total energy of charged supercells simulated via DFT and their relation to defect relaxation volumes are investigated in Sec. \ref{sec:IIIA}. Our results on the details of the lattice distortion around individual charged and neutral defects are presented in Sec. \ref{sec:IIIB}. In Sec. \ref{sec:IIIC}, we discuss the defect reaction approach for calculating the defect relaxation and formation volumes for complete defect formation reactions. In Sec. \ref{sec:IV}, we discuss the significance of our results and present our conclusions in Sec. \ref{sec:V}.

\section{\label{sec:II}Theoretical framework}
\subsection{\label{sec:IIA}Density functional theory}

Here, we summarize the details of the first-principles based DFT calculations used in this study. The DFT calculations are performed with the Vienna Ab Initio Simulation Package (VASP)\cite{Kresse1993,Kresse1996} using the projector-augmented-wave (PAW) method\cite{Blochl1994,Kresse1999}. For UO$_{2}$ the electron exchange and correlation potential is described by the local density approximation (LDA)\cite{Perdew1981} and the Hubbard+U model is used for the strongly-correlated U 5f electrons\cite{Liechtenstein1995}. In accordance with earlier studies, the U and J values are set to U=4.5 eV and J=0.51 eV; i.e. U$-$J=3.99 eV. We adopt the Perdew-Burke-Ernzerhof GGA (PBE-GGA) functional\cite{Perdew1996} for silicon and LDA for GaAs to be consistent with reference literature\cite{Kumagai2014, Freysoldt2011a}. The energy cut-offs employed are 500 eV for UO$_{2}$, 520 eV for Si, and 275 eV for GaAs. Defect properties in UO$_{2}$ are calculated using $2 \times 2 \times 2$ and $2 \times 2 \times 3$ supercells with $2 \times 2 \times 2$ and $2 \times 2 \times 1$ k-point meshes, respectively. Defect calculations are performed at both constant volume and fully relaxed (cell volume, shape, and atomic coordinates) conditions. Defects in silicon are simulated in a $2 \times 2 \times 2$ and $3 \times 3 \times 3$ size supercells with k-point meshes of $4 \times 4 \times 4$ and $3 \times 3 \times 3$, respectively, while those in GaAs are simulated in a $3 \times 3 \times 3$ supercell with a k-point mesh of $2 \times 2 \times 2$. The dielectric tensors are important for the correction schemes for the defect formation energies of charged systems and are calculated using density functional perturbation theory\cite{Baroni1986,Gajdos2006} as implemented in VASP. The computed bulk properties of interest for UO$_{2}$, Si, and GaAs are summarized in Table \ref{tab:1}.\par

\begin{table}[t]
	\caption{\label{tab:1} Calculated lattice parameter ($a_{0}$)
          in \AA, elastic constants ($C_{11}, C_{12}, C_{44}$), Bulk
          modulus (B) in GPa and (electronic+ionic) dielectric
          constant ($\varepsilon$) from DFT calculations in UO$_{2}$
          (LSDA+U) ($Fm\bar{3}m$), Si (GGA) ($Fd3m$) and GaAs (LDA)
          ($F\bar{4}3m$). Our results are indicated in bold, DFT and experimental literature values are provided for comparison.}
	\begin{ruledtabular}
		\begin{tabular}{ccccccc}
		 System & \multicolumn{1}{c}{$a_{0}$} & \multicolumn{1}{c}{$C_{11}$} & \multicolumn{1}{c}{$C_{12}$} & \multicolumn{1}{c}{$C_{44}$} & \multicolumn{1}{c}{B}   & \multicolumn{1}{c}{$\varepsilon$} \\[0.4em]
		\colrule \\ [-0.7em]
		\textbf{UO$_{2}$}					&5.451	&377.8	&133.2	&78.2	&220.5	&21.5	\\ [0.2em]
		DFT\footnote{Reference\citenum{Sanati2011}}		&5.448	&380.9	&140.4	&63.2	&220.6	&20.9	\\ [0.2em]
		Expt.	\footnote{Reference\citenum{Hampton1987a}}		&5.473	&389.3	&118.7	&59.7	&209.0	&23.6	\\ [0.2em]
		\textbf{Si}						&5.468		&153.1	&56.8	&74.6	&88.9	&13.1	\\ [0.2em]
		DFT\footnote{References\citenum{Lee2007, Baroni1986}}		&5.47	&154.6	&58.1	&74.4	&90.2	&12.9	\\ [0.2em]
		Expt.	\footnote{Reference\citenum{McSkimin1953, Madelung1991}}			&5.431	&167.7	&65.0	&80.4	&99.2	&11.2	\\ [0.2em]
		\textbf{GaAs}					&5.628	&117.4	&54.1	&56.0	&75.2	&18.7	\\ [0.2em]
		DFT\footnote{References\citenum{Kumagai2014, Komsa2012}}		&5.627	&124.2	&51.4	&63.4	&75.6	&17.9	\\ [0.2em]
		Expt.\footnote{Reference\citenum{Cooper1962, Garland1962}}				&5.654	&122.1	&56.6	&59.9	&76.9	&13.1	\\
		\end{tabular}
	\end{ruledtabular}
\end{table}

We perform convergence tests of the final total energy with respect to the energy cut-off and k-point mesh sizes. The convergence tolerance for forces on each ion is 0.01 eV/\AA, resulting in total energies converged to less than 10$^{-5}$ eV. To ensure there are no issues associated with the change in plane wave basis and k-point density due to changing cell size, we also calculate the defect relaxation volume of point defects for higher energy cut-off and denser k-point mesh; we find that the relaxation volumes of neutral and charged defects are converged to 0.3 - 1.8 \AA$^{3}$ (which is within 5\% of the relaxation volume) depending on the type of defect. Energy vs. volume calculations of defects are performed starting from the converged wave functions for the minimized defect structure. This is done to avoid the system becoming stuck in the metastable energy state associated with the possible configurations of the localized uranium f-electrons\cite{Dorado2013}.\par

Finally, to verify that our results are consistent across DFT codes, we performed calculations of the relaxation volume for defects in GaAs and Si with CASTEP\cite{Clark2005}, again using GGA-PBE with ``on the fly'' ultrasoft pseudopotentials with a cutoff energy of 450 eV. To ensure similarity with VASP simulations the same k-point grids are employed. The defect relaxation volumes as determined with CASTEP agree with those found with VASP, indicating that the behavior described here is not a consequence of a particular implementation of DFT.

\subsection{\label{sec:IIB}Defect formation and relaxation volumes}

The formation volume of a defect $\Delta\Omega^{f}$ is given by the change in Gibbs free energy, $\Delta G^{f} (P, T)$, for the defect formation reaction at constant pressure and temperature via\cite{Gillan1981}

\begin{equation}\label{eq:2}
\Delta G^{f} (P,T) = \Delta E^{f} (P,T) + P\Delta\Omega^{f} - T\Delta S(P,T).
\end{equation}
Here, $\Delta E^{f} (P,T)$ denotes the change in the internal energy, $\Delta\Omega^{f}$ the formation volume, and $\Delta S$ the change in entropy due to the formation of the defect at constant pressure and temperature. The defect formation volume is given as the difference in volume across the defect formation reaction and is related to the Gibbs free energy of defect formation by

\begin{equation}\label{eq:3}
\Delta \Omega^{f} = \Big( \frac{\partial \Delta G^{f} (P,T)}{\partial P} \Big)_{T, n_{i}}.
\end{equation}
At T = 0 K, the change in Gibbs free energy, $\Delta G^{f} (P, 0)$ reduces to the change in enthalpy, $\Delta H^{f}$. However, at higher temperatures, the entropy contribution can also become relevant\cite{Freysoldt2014}. The defect formation enthalpy, $\Delta H^{f}$, from DFT calculations for a defect X, in charge state q is given as\cite{Freysoldt2009, Freysoldt2014}

\begin{align}\label{eq:4}
\Delta H^{f} (X,q) &= \Big[ E_{\mathrm{tot}}^{\mathrm{DFT}} (X, q) + E_{\mathrm{corr}}^{\mathrm{Image}} +q\Delta V \Big] \\\nonumber 
 &- E_{\mathrm{tot}}^{\mathrm{DFT}} (\mathrm{bulk}) - \sum_{i}n_{i}\mu_{i} + q(\epsilon_{\mathrm{V}} + \epsilon_{F}) \\\nonumber 
 &+ P\Delta \Omega^{f},
\end{align}
where, $E_{\mathrm{tot}}^{\mathrm{DFT}} (X, q)$ is the total energy of the DFT supercell with the charged defect, $E_{\mathrm{tot}}^{\mathrm{DFT}} (\mathrm{bulk})$ is the total energy of the defect-free bulk supercell, and $n_{i}$ and $\mu_{i}$ are the number and chemical potential of each element contributing to the defect, $\epsilon_{\mathrm{V}}$ is the energy of the valence band maximum (VBM) calculated in the defect free bulk supercell\cite{}, and $\epsilon_{\mathrm{F}}$ is the Fermi energy with respect to the VBM. The term $q(\epsilon_{\mathrm{V}} + \epsilon_{F})$ is the chemical potential of electrons, i.e., it accounts for exchange of electrons with the reservoir. The $\Delta \Omega^{f}$ contribution to the enthalpy in Eq. \ref{eq:4}, is usually small at ambient conditions ($\sim$ 10$^{-5}$ eV per defect at P = 1atm) and often neglected. The terms $E_{\mathrm{corr}}^{\mathrm{Image}}$ and $\Delta V$ are the charged image interaction and potential alignment corrections, respectively. These corrections are needed to correct for finite-size artifacts within the DFT supercell approach to calculate the defect formation energies.\par

The expression for computing formation volume from DFT calculations is often given as\cite{Garikipati2006, Puchala2008}

\begin{align}\label{eq:5}
\Delta \Omega^{f} (X,q) &= \Delta \Omega^{r} (X,q) \pm \Omega_{0},
\end{align}
where,
\begin{align}\label{eq:6}
\quad \Delta \Omega^{r} &= \Omega (\mathrm{defect}) - \Omega (\mathrm{bulk}).
\end{align}
Here, $\Delta \Omega^{r}$ is the relaxation volume, and $\Omega (\mathrm{defect})$ and $\Omega (\mathrm{bulk})$ are the volumes of the supercell with a defect X in charge state q, and the perfect bulk supercell, respectively, both relaxed to zero stress. The + and $-$ signs in Eq. \ref{eq:5} denote vacancy and interstitial defects, respectively. The defect relaxation volume can be understood as the change in volume due to relaxation of the crystal lattice around the defect site, or the direct volume change in the system upon introducing the defect in a constant pressure calculation. By contrast, the formation volume also accounts for the reference volume (or the choice of reservoir) associated with the origin or placement of the species comprising the defect, $\Omega_{0}$. The relaxation, and formation volume are strictly a tensor quantities, e.g., $\Delta \Omega_{ij}^{r}$ is given as\cite{Puchala2008}

\begin{equation}\label{eq:7}
\Delta \Omega_{ij}^{r} = \Omega_{\mathrm{bulk}} \big[ \varepsilon_{ij}\big],
\end{equation}
where $\varepsilon_{ij}$ is the strain tensor (strain between supercells computed from DFT calculations) due to the defect. This definition holds for any anisotropy, shape or size of the supercell. The magnitude of the volume change is then given as, $\Delta \Omega^{r} = \frac{1}{3}\mathrm{tr}(\Delta \Omega_{ij}^{r})$, and is consistent with that computed using Eq. \ref{eq:6}. This description of the relaxation volume accounts for changes in cell shape upon introduction of the defect. It is important to note that relaxation volume approaches a finite, nonzero tensor for very large simulation cell size. Finally, the elastic dipole tensor $G_{ij}$ is related to $\Delta \Omega_{ij}^{r}$ (Eq. \ref{eq:1}) and thus, to describe the response of charged defects to strain, it is imperative to have accurate calculations of the relaxation volumes.

\subsection{\label{sec:IIC}Volume of individual species comprising the defect}

As described in Eq. \ref{eq:5}, the computation of the defect formation volume needs to take into account the reference volume, $\Omega_{0}$, of the individual atomic species constituting the defect reaction. For a single component system like silicon, $\Omega_{0}$ is equal to the cohesive volume of an atom in bulk silicon. However, for a multicomponent system, $\Omega_{0}$ is given by the partial molar volume, which measures the variation of volume as chemical components are added to or removed from the reservoir material during the defect reaction. The use of the reference material for the partial molar volume is equivalent to the use of the atomic chemical potential of the atomic species in the definition of the defect formation energy. In a multicomponent system defect reactions can be of two types:

(1) \textit{Stoichiometric reactions} comprising of defects like Schottky defects. In this case, $\Omega_{0}$ is the volume per formula unit of the multicomponent system, or, analogous to the monoatomic case, the cohesive volume of a formula unit. For example, in UO$_{2}$ this is computed to be $\Omega_{\mathrm{UO_{2}}} = 40.3$ \AA$^{3}$ (40.9 \AA$^{3}$ in experiment\cite{Desgranges2009}) and in GaAs it is $\Omega_{\mathrm{GaAs}} = 44.5$ \AA$^{3}$;\par

(2) \textit{Non-stoichiometric reactions} comprising of defects such as vacancies and interstitials. Here, $\Omega_{0}$, is the volume per unit of the reference. For example, in GaAs where solid bulk Ga and As can be considered as the reference material (or atomic reservoir) for respective defects, the computed volumes per unit are $\Omega_{\mathrm{Ga}} = 20.3$ \AA$^{3}$ (19.3 \AA$^{3}$ in experiment\cite{Heine1968}) and $\Omega_{\mathrm{As}} = 22.4$ \AA$^{3}$ (21.3 \AA$^{3}$ in experiment\cite{Schiferl1969}), respectively. That is, a reference is defined for each species that depends on the experimental conditions of interest. In this example, we chose bulk solids for each element. However, other reference states are possible or desired depending on the materials system and what experimental conditions are being matched. For example, in UO$_{2}$, bulk U metal in the orthorhombic structure ($\alpha$-phase) is considered as a reference for U vacancy and interstitial, with a computed volume of 19.9 \AA$^3$ (20.6 \AA$^3$ in experiment\cite{Barrett1963}). For O defects, the O$_{2}$ molecule is often taken as the reference state for computing defect formation energies. However, the computed volume of an oxygen molecule from ideal gas law is $\sim$10$^{5-6}$ \AA$^{3}$ for a given pressure and temperature; thus O$_{2}$ gas may not make sense as a reference material for the defect volume. We have therefore adopted a slightly different approach by choosing orthorhombic $\alpha$-U$_3$O$_8$ as the reference for both chemical potentials and partial molar volumes for U vacancy and O interstitial in UO$_2$, with a computed volume of 56.9 \AA$^3$ per U atom compared to 55.5 \AA$^3$ per U atom from experiment\cite{Allen1995}. In all of these cases, the reference explicitly depends on the experimental conditions and is not something directly computable from electronic structure methods. That is, once a reference is chosen, the relevant thermodynamic properties can be calculated, but the reference itself is not something that can be determined solely from the calculations: it must be asserted or assumed.

We use Eqs. \ref{eq:5} and \ref{eq:6} to calculate the defect formation and relaxation volumes. In compounds this calculation is complicated by the fact that the stoichiometry of the system changes when individual point defects such as vacancies or interstitials are formed. For charged defects, the issue is further exacerbated as the electrons or holes added to the defect system also contribute to both the defect formation energy and volume. While the contribution to the defect formation energy has been well studied and several correction terms have been developed, the contribution of the change in charge to the defect formation volume has not been considered systematically.\par

In the following discussion, we explicitly focus on the defect relaxation rather than the formation volume as the latter can be easily be obtained by properly accounting for $\Omega_{0}$. Here we show that the contribution of the change in charge to defect relaxation volumes is surprisingly large. By studying point defects in three different systems, we estimate for a given system contribution of the charge to the defect relaxation volume.\par

\section{\label{sec:III}Results}

\begin{table}[t]
	\caption{\label{tab:2} Computed defect relaxation volumes,
          $\Delta \Omega^{r}$ (\AA$^{3}$), normalized relaxation volume,
          $\Delta \Omega^{r}/\Omega_{\mathrm{unit}}$ for charged and neutral point defects
          calculated from DFT calculations. For reference, for UO$_{2}$ and GaAs,
          the volume per formula unit ($\Omega_{\mathrm{unit}}$) is 40.3 \AA$^{3}$ ($\Omega_{\mathrm{UO_{2}}}$),
          and 44.6 \AA$^{3}$ ($\Omega_{\mathrm{GaAs}}$), respectively, and for Si the atomic volume is 
          20.4 \AA$^{3}$ ($\Omega_{\mathrm{Si}}$). The values are obtained
          from 144 atom UO$_{2}$ supercell, and 216 atom supercell for Si and GaAs.}
		\begin{ruledtabular}
		\begin{tabular}{llddd}
		System & Defect & \multicolumn{1}{c}{$\Delta \Omega^{r}$} & \multicolumn{1}{c}{$\Delta \Omega^{r}/\Omega_{\mathrm{unit}}$}\\[0.3em]
		\colrule \\ [-0.5em]
		UO$_{2}$	&V$^{0}_{\mathrm{U}}$			&-9.1	&-0.22\\ [0.4em]
				&V$^{4-}_{\mathrm{U}}$ 			&40.5	&1.05\\ [0.4em]
				&V$^{0}_{\mathrm{O}}$ 			&0.56	&0.01\\ [0.4em]
				&V$^{2+}_{\mathrm{O}}$ 			&-16.7	&-0.41\\ [0.4em]
				&O$^{0}_{i}$ 			&-3.38	&-0.08\\ [0.4em]
				&O$^{2-}_{i}$ 			&20.0	&0.49\\[0.4em]
		GaAs	&V$_{\mathrm{Ga}}^{0}$			&-26.3	&-1.12\\ [0.4em]
				&V$_{\mathrm{Ga}}^{3-}$		&31.2	&1.42\\ [0.4em]
				&V$_{\mathrm{As}}^{0}$			&-24.4	&-1.09\\ [0.4em]
				&V$_{\mathrm{As}}^{3+}$		&-73.5	&-3.3\\ [0.4em]
		Si		&V$_{\mathrm{Si}}^{0}$			&-14.7	&-0.72\\ [0.4em]
				&V$_{\mathrm{Si}}^{2+}$			&-42.8	&-2.1\\ [0.4em]
				&V$_{\mathrm{Si}}^{2-}$			&28.1	&1.38\\ 
		\end{tabular}
	\end{ruledtabular}
\end{table}

\begin{figure*}[t]
\includegraphics[width=0.8\textwidth]{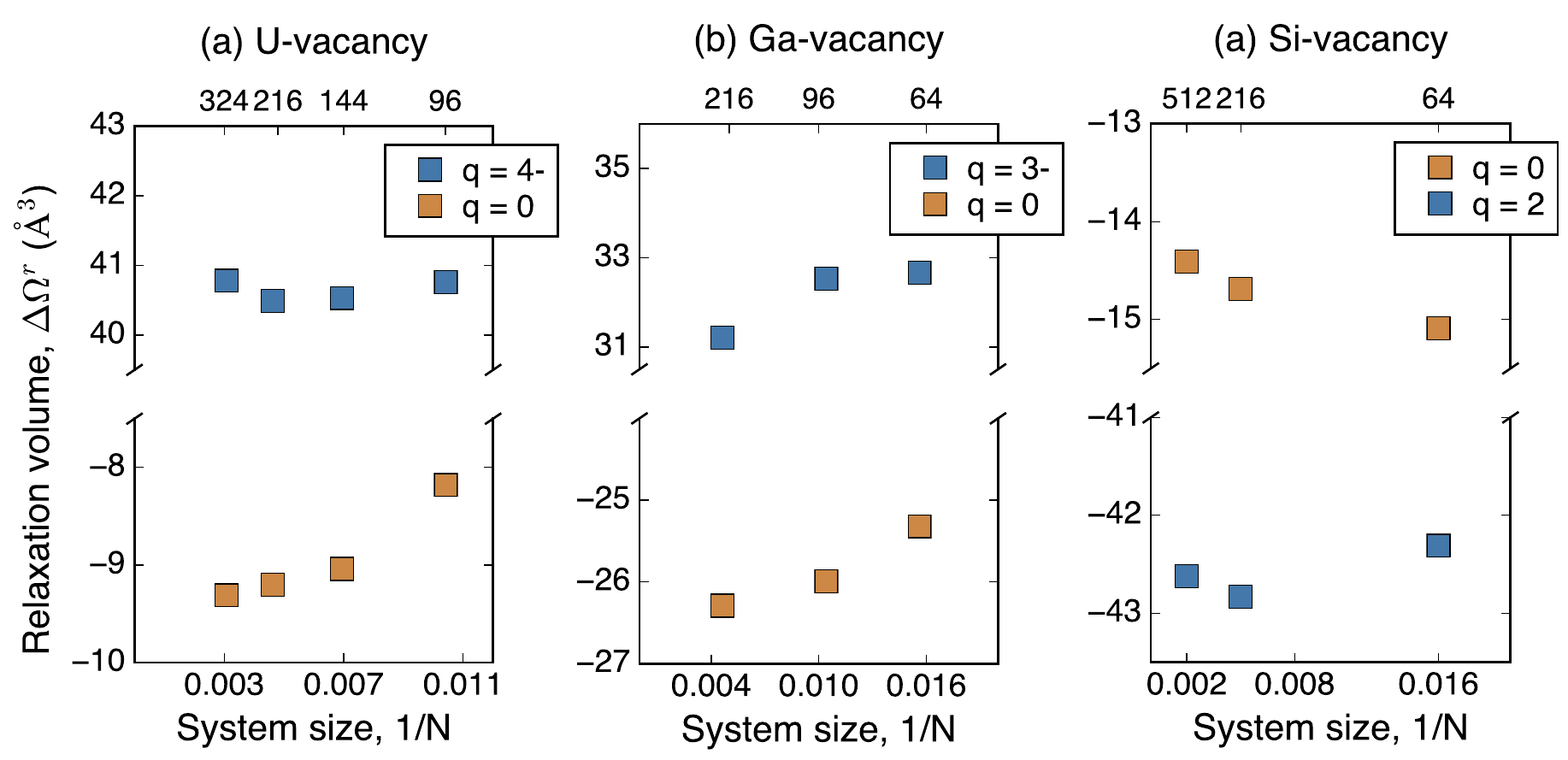}
\caption{\label{fig:1}Relaxation volume of a single neutral and charged (a) uranium vacancy in UO$_{2}$, (b) Ga vacancy in GaAs, and (c) Si vacancy as function of the inverse of the supercell size (the corresponding number of atoms in the supercell are provided on the upper x-axis).}
\end{figure*}

The computed values of the relaxation volumes of point defects in both neutral and fully charged states are listed in Table \ref{tab:2}. The full estimation of lattice relaxation and symmetry around a defect is quantified by the relaxation volume tensor, $\Delta \Omega_{ij}^{r}$ (Eq. \ref{eq:7}), which in turn determines the defect dipole tensor, $G_{ij}$ (Eq. \ref{eq:1}), listed in Table \ref{tab:3}. The relaxation volumes are relatively large for charged defects, compared to their neutral counterpart, and, interestingly, the sign of the relaxation volume is sensitive to the charge state of the defect. The relaxation volumes are computed using Eq. \ref{eq:6} (when relaxing cell shape, volume and ionic positions), and Eq. \ref{eq:7} (when keeping volume and cell shape fix, relaxing only ionic positions). The agreement in computed relaxation volumes calculated through the two approaches is between 0.5-2 \AA$^3$; the differences in total energies is 50-100 meV, a similar range to that reported by Freysoldt et al.\cite{Freysoldt2014}.\par

As discussed above, the dipole tensor is used to compute the change in defect formation energy upon interaction with strains in the material\cite{Hinterberg2013, Goyal2015}. We find that the changes in the formation energy differ significantly between the charged and neutral states for a given defect. For example, for a fully charged (2-) oxygen interstitial under 1\% compressive strain the change in formation energy is about 277.4 meV, which is 6 times larger than the $-$46.4 meV change in energy for a neutral oxygen interstitial under the same strain state. Similarly, the segregation energy of a fully charged oxygen interstitial in UO$_{2}$ under the strain field of edge dislocation\cite{Goyal2013} is computed to be about $-$8.3 eV, much larger than the $-$1.2 eV segregation energy of the neutral oxygen interstitial. This extremely large change in energy between a charged and a neutral defect interacting with a strain field is unphysical and suggests that something is amiss with the volumes calculated for the charged defects.\par

Figure \ref{fig:1} shows the dependence of the defect relaxation volume on the simulation cell size for charged and neutral vacancies. These results thus indicate that the large charged defect relaxation volumes are not a consequence of the finite system size limitation of DFT.  In the next section, we investigate in detail the dependence of relaxation volumes on the various finite size charge corrections schemes to charge supercell calculations, which have been suspected\cite{Bruneval2012} to have large volume dependence.\par

\begin{table}[b]
	\caption{\label{tab:3} Computed defect dipole tensors, for fully charged and neutral point defects calculated from DFT calculations.}
		\begin{ruledtabular}
		\begin{tabular}{lldddddd}
		Defect & Charge & \multicolumn{1}{c}{$G_{11}$} & \multicolumn{1}{c}{$G_{22}$} & \multicolumn{1}{c}{$G_{33}$} &
		\multicolumn{1}{c}{$G_{12}$} & \multicolumn{1}{c}{$G_{13}$} & \multicolumn{1}{c}{$G_{23}$}\\[0.2em]
		\colrule \\ [-0.7em]
		V$_{\mathrm{U}}$	&4-	&56.08	&55.18	&56.97	&0.32	&0.19	&0.11\\ [0.3em]
						&0	&-10.52	&-16.86	&-9.04	&-0.16	&3.06	&-1.48\\ [0.3em]
		V$_{\mathrm{U}}$	&2+	&-22.12	&-24.01	&-22.69	&-0.14	&0.02	&0.05\\ [0.3em]
						&0	&0.24	&1.02	&1.06	&0.02	&-0.01	&0.01\\ [0.3em]
		O$_{i}$			&2-	&27.87	&26.81	&28.56	&-0.88	&-0.37	&-0.33\\ [0.3em]
						&0	&-4.83	&-4.82	&-4.29	&-0.32	&0.15	&0.01\\
		\end{tabular}
	\end{ruledtabular}
\end{table}

\subsection{\label{sec:IIIA}Finite size correction schemes to energy for charged supercell}

As discussed in the Sec. \ref{sec:IIB}, the defect formation volume follows thermodynamically from the defect formation energy (Eq. \ref{eq:3}). It is therefore important to investigate any possible connection between the corrections to the defect formation energy of charged defects and their defect formation volumes.\par

Charged supercell calculations in DFT require an artificial background charge density (opposite in sign to the net charge of the supercell) to ensure global charge neutrality and convergence of the total energy of the charged system within the supercell approach. The genesis of this background charge is related to the DFT convention of setting the average electrostatic potential over the entire volume of the supercell to zero in order to evaluate the electrostatic Coulomb energy\cite{Makov1995}. We note that the same convention is typically applied to calculations of charged defects using ionic potentials. The total energy of charged supercells must account for the chemical potential of the electrons, $q(\epsilon_{\mathrm{V}} + \epsilon_{F})$ in Eq. \ref{eq:4}, and needs to be corrected for finite size errors associated with the periodic supercell cell approach\cite{Makov1995, Lany2008, Freysoldt2011}. There are two specific effects. The first is the interaction between a charged defect and its periodically repeated charged images, adding a spurious electrostatic energy; this is referred to as the image charge correction, $E_{\mathrm{corr}}^{\mathrm{Image}}$. The second is the electrostatic potential between the defect and bulk system that needs to be aligned in order to accurately define the reference energy (chemical potential) of electrons, added or removed in charged supercell calculations. This is termed as the potential alignment correction, $q\Delta V$. We adopt the method by Freysoldt et al.\cite{Freysoldt2011} to correct for finite size errors as it has been shown to achieve good convergence\cite{Komsa2012} for defect formation energies with system size.\par

In the next subsections, we systematically investigate the relation between these finite-size energy corrections and the charged defect relaxation volumes. Any change in total energy that depends on the volume of the simulation cell affects the computed forces, pressure/stress tensors, and relaxation volumes in DFT simulations. Therefore, both volume and stresses need to be corrected in the same manner as the total energy.\par
								
\subsubsection{\label{sec:IIIA1}Image charge correction}

The image charge correction is needed because, in periodic supercell calculations, a charged defect (in the presence of the homogeneous background charge density) has long-range electrostatic interactions with its periodic images. If the dimensions of the unit cell were infinite, this electrostatic interaction would be zero. However, in periodic DFT calculations the supercell dimensions are finite; therefore we need to correct the total energy of the charged system to account for this spurious Coulomb interaction. This correction is referred to as the monopole correction by Leslie and Gillan\cite{Leslie1985}, and is designed to transform the electrostatic energy of a periodic lattice of point charges in a neutralizing background into the electrostatic energy of single point charge. As such, it is not unique to DFT, but occurs whenever the energetics of a system with net charge is considered. The correction term is given as\cite{Leslie1985, Freysoldt2011}

\begin{table}[t]
	\caption{\label{tab:4} Calculated energy correction ($\Delta E_{\mathrm{corr}}^{\mathrm{Image}}$) to the defect formation energy (in eV) due to image interaction between charged defects, and the corresponding correction to pressure (in GPa) (1 eV/\AA$^{3}$ = 160.2 GPa) and volume (in \AA$^{3}$) for representative charged defects in 144 atoms UO$_{2}$ supercell, and 216 atoms supercell for Si and GaAs.}
		\begin{ruledtabular}
		\begin{tabular}{ldddd}
		Defect & \multicolumn{1}{c}{$\Delta E_{\mathrm{corr}}^{\mathrm{Image}}$} & \multicolumn{1}{c}{$\Delta P_{\mathrm{corr}}^{\mathrm{Image}}$} 
		& \multicolumn{1}{c}{$\Delta \Omega_{\mathrm{corr}}^{\mathrm{Image}}$} & \multicolumn{1}{c}{$\Delta \Omega^{r}$}\\[0.2em]
		\colrule \\ [-0.7em]
		V$_{\mathrm{U}}^{4-}$	&1.07	&-0.03	&0.28	&40.2	\\ [0.3em]
		V$_{\mathrm{Ga}}^{3-}$	&0.55	&-0.01	&0.39	&31.2	\\ [0.3em]
		V$_{\mathrm{Si}}^{2-}$	&0.57	&-0.02	&0.34	&16.5	\\
		\end{tabular}
	\end{ruledtabular}
\end{table}

\begin{equation}\label{eq:8}
\Delta E_{\mathrm{corr}}^{\mathrm{Image}} = E_{\mathrm{isolated}} - E_{\mathrm{periodic}} = \frac{\alpha q^{2}}{2\varepsilon \Omega^{1/3}}.
\end{equation}
Here, $\alpha$ is the Madelung constant, $\Omega$ is the volume of the bulk supercell in which the defect is created, and $\varepsilon$ represents the electronic and ionic low frequency dielectric constant of the material, computed from DFT calculations (Table \ref{tab:1}). In our calculations, we use the average value\cite{Murphy2013}, $\frac{1}{3}\mathrm{tr}(\varepsilon_{ij})$ of the dielectric tensor as the effective value for dielectric constant in Eq. \ref{eq:8}.\par

The energy correction given by Eq. \ref{eq:8} is computed directly from VASP, and the corresponding correction to the pressure $\Delta P_{\mathrm{corr}}^{\mathrm{Image}}$ can either be calculated directly from the slope of the image correction ($\Delta E_{\mathrm{corr}}^{\mathrm{Image}}$) as a function of volume or via the relation

\begin{equation}\label{eq:9}
\Delta P_{\mathrm{corr}}^{\mathrm{Image}} = -\frac{\partial E_{\mathrm{corr}}^{\mathrm{Image}}}{\partial \Omega}
= \frac{1}{3} \Big( \frac{\alpha q^{2}}{2\varepsilon \Omega^{4/3}} \Big)
\end{equation}
with the corresponding volume correction given as,

\begin{equation}\label{eq:10}
\Delta \Omega_{\mathrm{corr}}^{\mathrm{Image}} = -\Omega_{\mathrm{Bulk}} \Big( \frac{\Delta P_{\mathrm{corr}}^{\mathrm{Image}}}{B} \Big).
\end{equation}

\begin{figure}[t]
\includegraphics[width=0.425\textwidth]{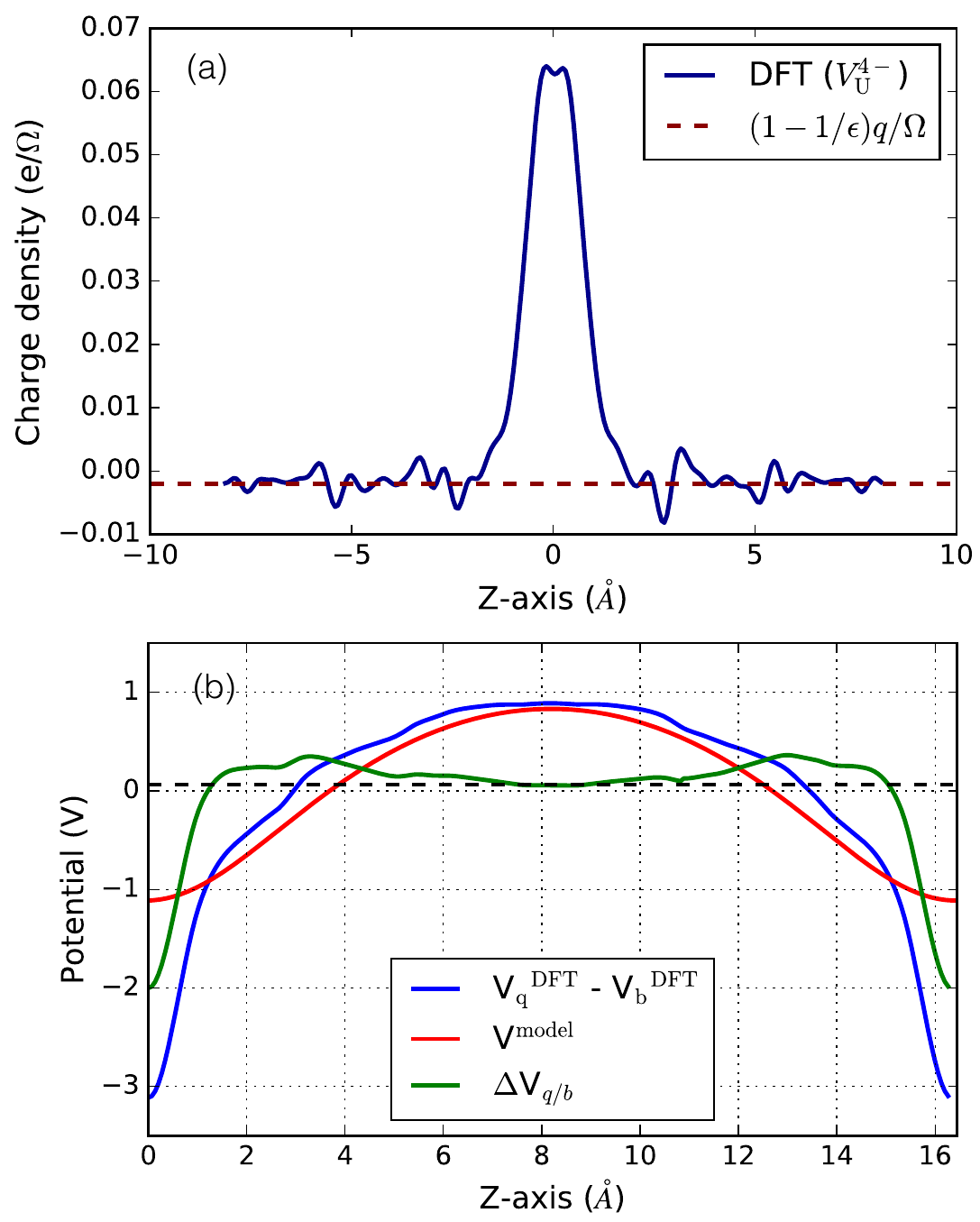}
\caption{\label{fig:2} (a) Planar-averaged charge density difference across the supercell for V$_{\mathrm{U}}^{4-}$ as obtained from static (non-relaxed) DFT calculations. The difference is taken between the charge densities of the charged defect and bulk. (b) Planar-averaged potential difference ($V_{\mathrm{q}}^{\mathrm{DFT}} - V_{\mathrm{b}}^{\mathrm{DFT}}$) across the supercell for V$_{\mathrm{U}}^{4-}$, as obtained from DFT calculations (blue curve). The model potential ($V^{\mathrm{model}}$) from the Gaussian charge is shown in red, and the difference potential $\Delta V_{\mathrm{q/b}}$ is shown in green. The z-axis distance is measured from the location of the defect. The potential falls off again at a distance of $\sim$16\AA, as that is the size of the cell and where the periodic image of the defect is located.}
\end{figure}

The computed image charge corrections to energy, pressure and volume are summarized in Table \ref{tab:4} for charged defects in UO$_{2}$, GaAs and Si. The corrections to volume are very small compared to the corrections in the total energy and to the corresponding defect relaxation volumes: $\Delta \Omega_{\mathrm{corr}}^{\mathrm{Image}} $ $\sim$ 0.01-0.02 $\Delta \Omega^{r}$, for the three systems studied here. Thus, the correction to the relaxation volumes due to the image charges is only a 1-2\% effect.

Our computed value of $\Delta \Omega_{\mathrm{corr}}^{\mathrm{Image}}$ for the charged (2+) Si vacancy is 0.02$\Omega_{\mathrm{Si}}$ , which is similar to that reported by Bruneval and co-workers\cite{Bruneval2012}. We note that the correction to the relaxation volume and pressure (or stresses) decreases with the increasing size of the simulation supercell and is independent of the sign of the charge on the defect, depending only on the magnitude of the charge. For example, for the U vacancy, the energy and volume corrections are 0.82 eV, and 0.2 \AA$^{3}$ in a $3 \times3 \times3$ supercell, compared to 1.23 eV, and 0.32 Å3 in $2 \times2 \times2$ supercell.\par

The above changes in the pressure and volume are computed based on two assumptions. The first assumption is that the value of the bulk modulus, B, does not change significantly due to the presence of defects. The second assumption is that the charge is localized at the defect site within the simulated supercell size. Our computed value of the bulk modulus of UO$_{2}$ with a U vacancy (in a 144 atom supercell) is 203.3 GPa compared to the bulk value of 220.5 GPa.This difference gives an error of 0.01-0.03 \AA$^{3}$ in the estimation of $\Delta \Omega_{\mathrm{corr}}^{\mathrm{Image}}$. We also computed the difference in planar averaged charge density of the charged defect (averaged over planes of atoms) from its bulk structure, as shown in Fig. \ref{fig:2}(a), for the charged uranium vacancy in UO$_{2}$. The charge density is fairly localized at the defect site and saturates to the screened54 charge value of $\big(1-\frac{1}{\varepsilon}\big)\frac{q}{\Omega}$, as indicated by dashed red line in Fig. \ref{fig:2}(a).\par

\subsubsection{\label{sec:IIIA2}Alignment correction}

Among the various procedures\cite{Lany2008, Freysoldt2011, Taylor2011} outlined in the literature to align the average electrostatic potential in the defect supercell with that in the bulk supercell, we use the scheme provided by Freysoldt et al.\cite{Freysoldt2011}. The key advantage of this scheme (referred to in the literature as the FNV scheme\cite{Komsa2012}) is that the long-range 1/r potential is removed before the alignment is determined for the remaining short-range potential. If the size of the simulation cell is big enough to fully localize the defect, the short range potential reaches a plateau (Fig. \ref{fig:2}(b)) far from the defect, which yields the alignment as a well-defined quantity\cite{Freysoldt2014}.\par

\begin{figure}[b]
\includegraphics[width=0.49\textwidth]{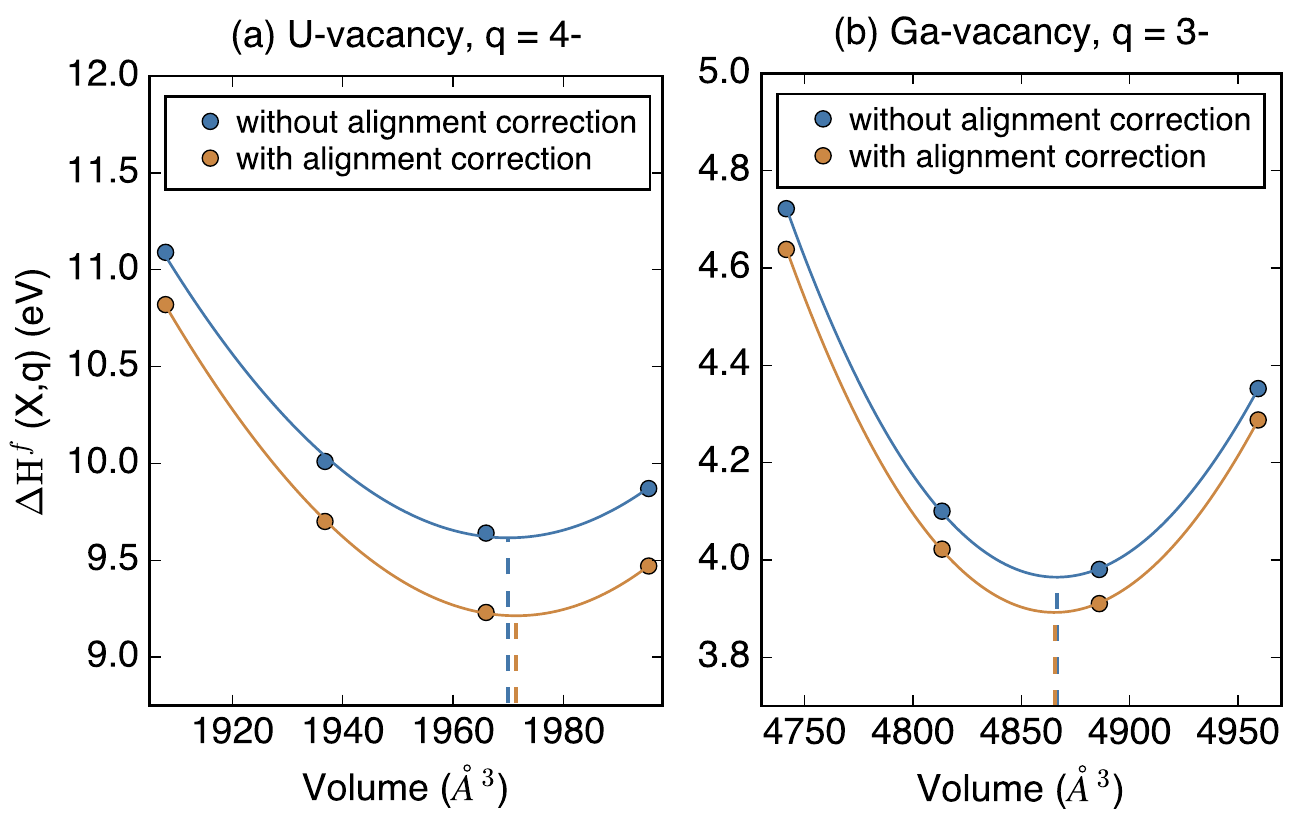}
\caption{\label{fig:3} Formation energy of charged defects with and without the alignment correction for (a) $\Delta V$ for V$_{\mathrm{U}}^{4-}$, (b) V$_{\mathrm{Ga}}^{3-}$ as function of scaled volume of defect supercell. The corresponding relaxation volumes extracted from these calculations are indicated by the dashed lines.}
\end{figure}

The short-range potential away from the defect site is obtained by taking the difference between the electrostatic potential from a model charge density $V^{\mathrm{model}}$ and the DFT electrostatic potential from the charged defect $V_{\mathrm{q}}^{\mathrm{DFT}}$ and the perfect bulk $V_{\mathrm{b}}^{\mathrm{DFT}}$:

\begin{equation}\label{eq:11}
\Delta V = \Delta V_{q/b} = \big( V_{\mathrm{q}}^{\mathrm{DFT}} - V_{\mathrm{b}}^{\mathrm{DFT}} \big)|_{\mathrm{far}} - V^{\mathrm{model}}|_{\mathrm{far}} .
\end{equation}
The model charge density is the charge distribution used to capture the localized defect charge (as shown in Fig. \ref{fig:2}(a)). In the FNV alignment correction scheme, we use a Gaussian charge distribution for the model charge density, with width equal to 1 Bohr. The DFT electrostatic potential (blue curve), and the potential due to model charge density (red curve) are shown in Fig. \ref{fig:2}(b). The difference is the short-range electrostatic potential (green curve) which reaches a plateau away from the defect site (centered at origin), thereby giving the alignment correction $\Delta V$ (highlighted as dashed line in the plots). The computed value of $\Delta V$ for V$_{\mathrm{U}}^{4-}$ is 0.07 eV, and is computed in a similar manner for all other charged defect studied.\par

There is no well-defined formula for calculating the alignment correction; hence to obtain the corresponding corrections to pressure and volume, we compute the defect formation energy with and without the alignment correction for a set of isotropically-scaled volumes. Figure \ref{fig:3} shows the defect formation energy as function of volume; from the fit to the energy-volume data we compute the volume corresponding to the energy minima. Comparing the volumes found from the defect formation energies with and without the alignment correction gives an estimation of the correction to the relaxation volume due to the alignment correction. For V$_{\mathrm{Ga}}^{3-}$ in a 216 atom supercell, the volume correction is -0.91 \AA$^{3}$ (0.02$\Omega_{\mathrm{GaAs}}$), while that for $\Delta V$ for V$_{\mathrm{U}}^{4-}$ in 144 atom supercell is about 1.6 Å3 (0.039$\Omega_{\mathrm{UO_{2}}}$). Again, these corrections are small, on the order of 2-4\% of the total relaxation volume.\par

\begin{figure}[t]
\includegraphics[width=0.425\textwidth]{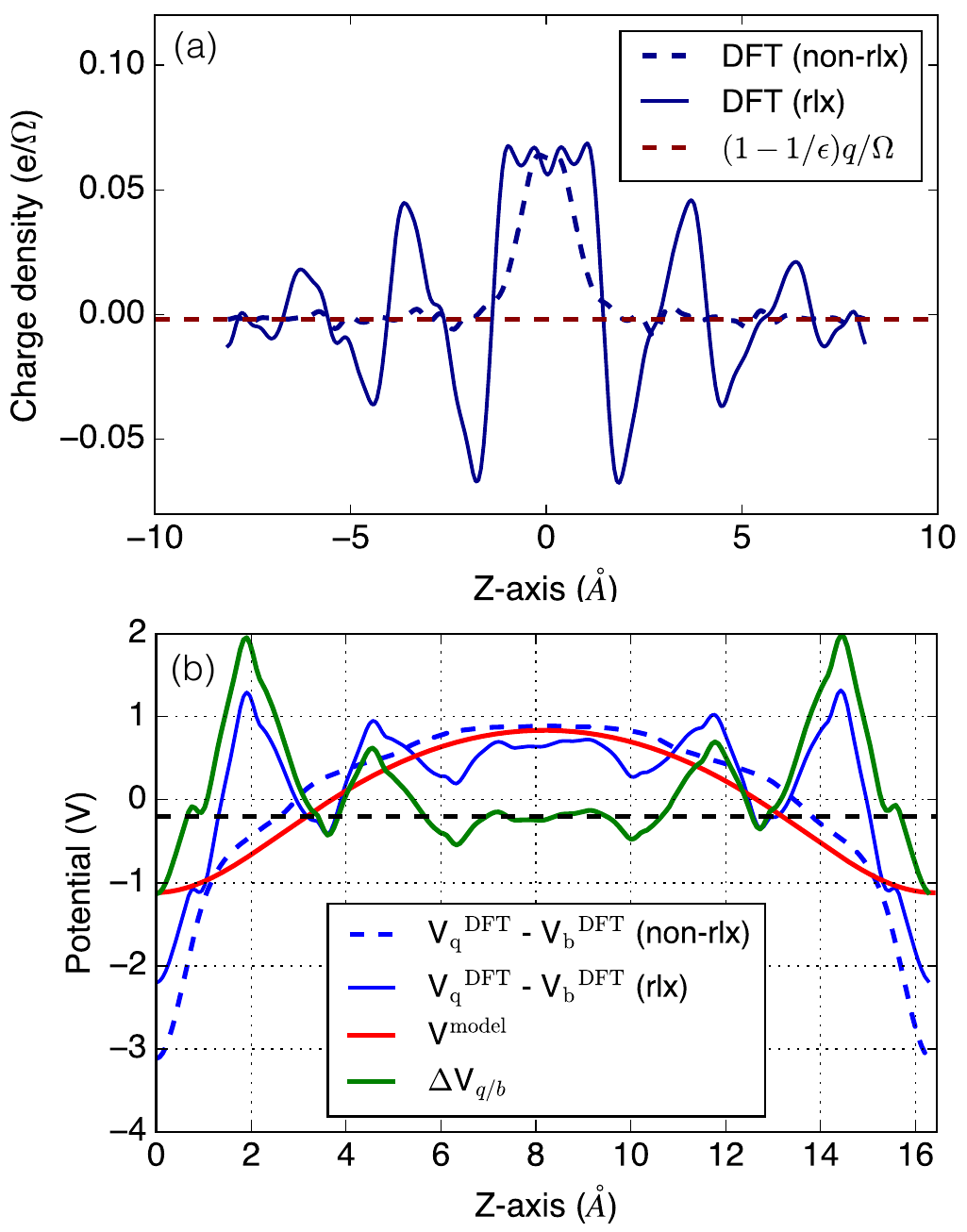}
\caption{\label{fig:4} Planar-averaged (a) charge density, and (b) potential difference plots, for V$_{\mathrm{U}}^{4-}$, showing results for both static (Fig. \ref{fig:2}) and relaxed DFT calculations.}
\end{figure}

These results so far suggest that both the image charge and alignment corrections to charged defects have very weak volume (or pressure) dependences. Both corrections are post-processing methods to correct the defect formation energy. For defects with large atomic relaxations, the charge density and potential can show strong variations close to the defect site, as shown for the uranium vacancy in Fig. \ref{fig:4}. This results in uncertainty in extracting the exact value of the alignment correction. Recent reviews\cite{Komsa2012, Kumagai2014} of charged defect correction schemes highlight similar issues with relaxed geometries. While this source of error in potential alignment correction needs to be quantified, the results from Ref. \citenum{Freysoldt2014} suggest that the changes are relatively small, i.e. within 40-100 meV, and unlikely to make significant change to its volume dependence.\par

\subsection{\label{sec:IIIB}Structure of charged defects}

In this section, we investigate in detail the atomic structure of charged defects, to gain insight into the differences in local lattice relaxation around the neutral and charged defects, and thus understand the origin of the anomalously large relaxation volumes of charged defects.

\begin{table}[t]
	\caption{\label{tab:5} Relaxation volume, $\Delta \Omega^{r}$ (\AA$^{3}$), of an electron and a hole from DFT calculations in Si, GaAs, and UO$_2$. All volumes are given relative to the volume of one formula unit in the perfect supercell. These are $\Omega_{\mathrm{Si}}$ = 20.4 \AA$^{3}$, $\Omega_{\mathrm{GaAs}}$ = 44.56 \AA$^{3}$, and $\Omega_{\mathrm{UO_{2}}}$ = 40.35 \AA$^{3}$.}
		\begin{ruledtabular}
		\begin{tabular}{llll}
		Defect &  Si & GaAs & UO$_2$ \\[0.2em]
		\colrule \\ [-0.7em]
		Electron	&0.74$\Omega_{\mathrm{Si}}$		&0.53$\Omega_{\mathrm{GaAs}}$	&0.28$\Omega_{\mathrm{UO_{2}}}$	\\ [0.3em]
		Hole		&-0.84$\Omega_{\mathrm{Si}}$	&-0.48$\Omega_{\mathrm{GaAs}}$	&-0.28$\Omega_{\mathrm{UO_{2}}}$	\\
		\end{tabular}
	\end{ruledtabular}
\end{table}

\begin{figure}[b]
\includegraphics[width=0.425\textwidth]{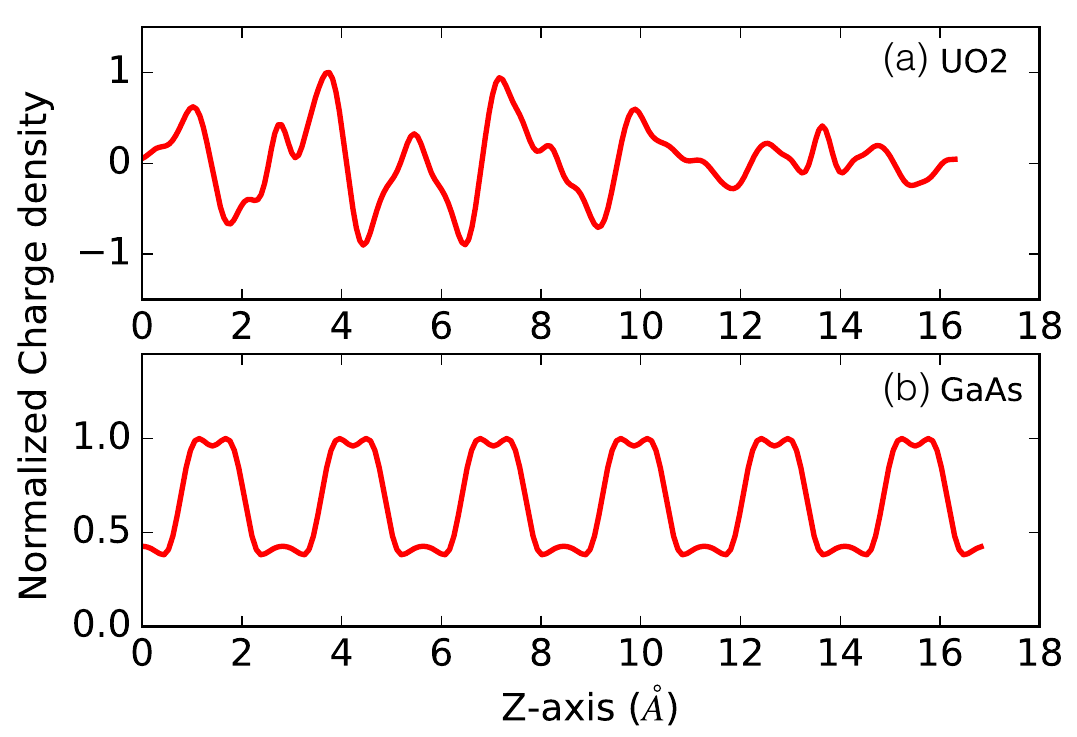}
\caption{\label{fig:5} Difference in charge density (xy-planar averaged) between supercells containing an extra electron and bulk along the z-axis in (a) UO$_2$, and (b) GaAs. The difference in charge density is normalized by the maximum value of the difference.}
\end{figure}

\begin{figure*}[t]
\includegraphics[width=0.85\textwidth]{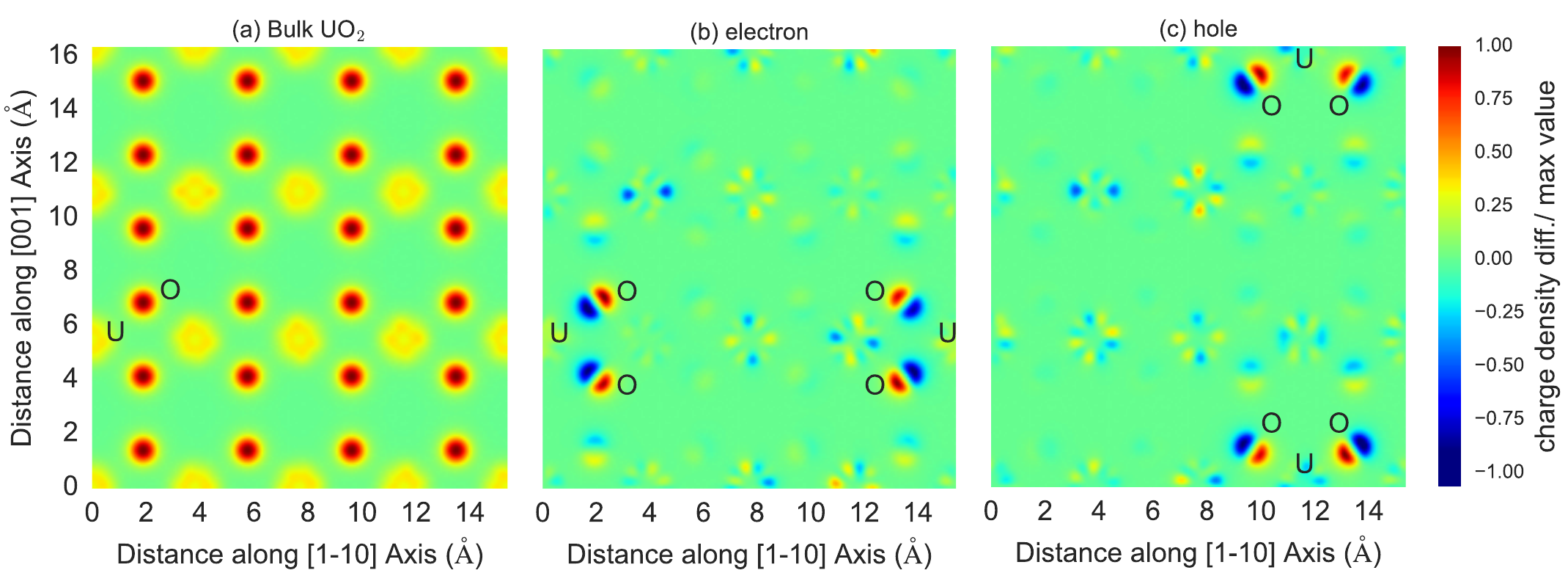}
\caption{\label{fig:6} (a) Charge density (slice along a (110) plane containing both U and O atoms) for bulk UO$_2$, (b) Difference in charge density (within the same (110) plane) between the supercell with the extra electron and bulk UO$_2$, and (c) between the supercell containing a hole and bulk UO$_2$.}
\end{figure*}

\subsubsection{\label{sec:IIIB1}Lattice relaxation around an electron and hole}

An electron or hole is simulated in DFT calculation by changing the total number of electrons by +1 or $-$1, in the optimized bulk structures. The relaxation volumes of electrons and holes in bulk Si, GaAs, and UO$_{2}$, obtained after relaxing (P=0) the charged bulk structures, are summarized in Table \ref{tab:5}. Addition and removal of an electron result in expansion and contraction of the lattice, respectively. These volume changes ($\sim$12 to 22 \AA$^{3}$ or 0.3 to 0.7$\Omega_{0}$, depending on the system) are large and result in about 1\% change in their respective lattice parameters for the current supercell sizes. There is not much data in the literature on the relaxation volumes of electrons and holes to compare our results. However, a few DFT studies\cite{Centoni2005, Windl2002} on defects in Si report the relaxation volumes of an electron of 0.68 to 0.72$\Omega_{\mathrm{Si}}$ and a hole of $-$0.78$\Omega_{\mathrm{Si}}$, where $\Omega_{\mathrm{Si}}$ is the volume per Si atom in a perfect supercell. Our computed values for the relaxation volume of an electron and hole in Si are in a similar range: 0.74$\Omega_{\mathrm{Si}}$ and $-$0.84$\Omega_{\mathrm{Si}}$, respectively.\par

A DFT calculation of an electron or hole often yields a delocalized electron, as it represents the exchange of electron or hole with the bulk conduction band or the valence band edge, unless the added (or removed) extra charge has a preference to localize on specific atomic site. To verify this, we performed Bader charge analysis\cite{Bader1990, Henkelman2006} and looked at the difference in charge density (planar averaged) between the electron (or hole) and bulk for both UO$_{2}$ and GaAs.\par

For GaAs, the difference in charge density (Fig. \ref{fig:5}) between the electron and bulk is uniformly distributed across the As and Ga atomic planes normal to the z-axis. However, for UO$_2$, the difference in charge density is not uniform (Fig. \ref{fig:5}), indicating localization of the additional electrons within the supercell. A 2D slice (Fig. \ref{fig:6}) of the difference in charge density between the excess electron and bulk shows large variations in charge density across specific U and O sites. Further, a Bader charge analysis on the structure with an extra electron in UO$_{2}$ confirms a U site having an effective Bader charge of +2.18, less than that of +2.54 on a U site in the bulk (or perfect) lattice, and neighboring O sites with Bader charge of $-$1.30, larger than of $-$1.27 on O site in bulk. A similar Bader charge analysis of the hole suggests partial localization at a U site with an effective Bader charge of +2.77. Such partial localization of an extra electron ($e^{'}$ or U$_{\mathrm{U}}^{\delta-}$) or hole ($h^{\cdot}$ or U$_{\mathrm{U}}^{\delta+}$) has previously been observed in DFT+U studies of UO$_{2}$\cite{Dorado2010}.\par

\begin{figure*}[t]
\includegraphics[width=0.8\textwidth]{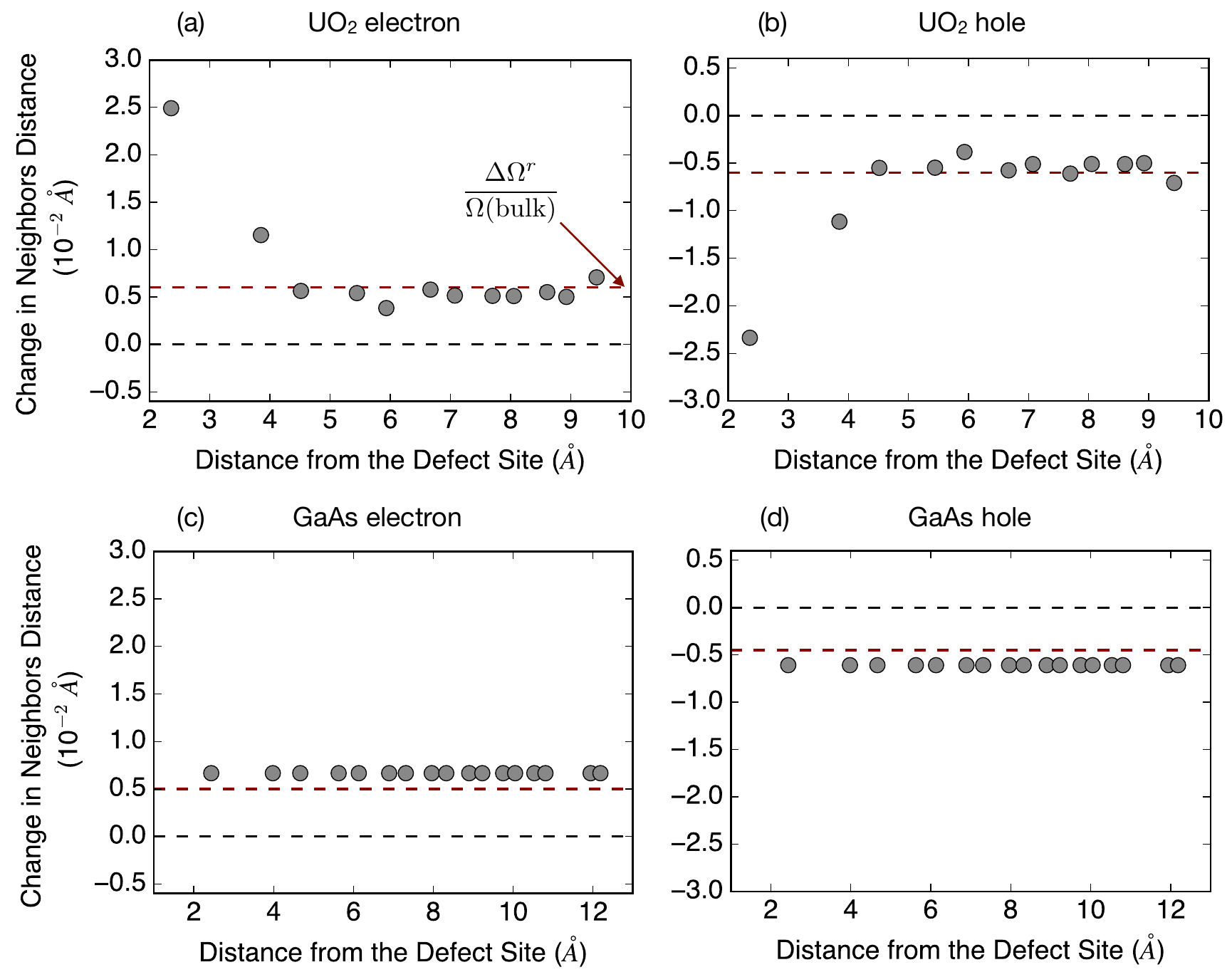}
\caption{\label{fig:7} Change in nearest neighbor distances (in percent) between the structures with an extra electron (or hole) and the bulk: (a) electron, and (b) hole in UO$_2$; (c) electron, and (d) hole in GaAs.}
\end{figure*}

To gauge the local atomic relaxations due to the presence of an excess electron or hole, we analyze the change in distances of the neighboring atoms for all the sites in the perfect and defective (electron or hole) structure. Figure \ref{fig:7} shows the average change in neighboring distances for all the sites in the defect structure, plotted as the function of distance from the defect site. Changes in distance up to 2nd nearest neighbors are considered for each site between the bulk and defect structure, and are normalized by the total number of neighboring atoms.\par

In UO$_{2}$ with a partially localized extra electron on a U site ($e^{'}$ or U$_{\mathrm{U}}^{\delta-}$)  (Fig. \ref{fig:7}(a)), the local defect structure is complex. Neighboring atoms for any given site in the structure show outward relaxation, independent of the position of the site from the defect. For those sites at a distance of $\sim$5 \AA, and beyond from the defect, the average change in neighboring distance saturates to a constant value. This constant value is approximately equal to the fractional volume change ($\Delta \Omega^{r}/\Omega_{\mathrm{bulk}}$ =11.52/1936.76) due to the addition of an electron in bulk UO$_2$, which is about $\sim$ 0.006 (or 0.6\%). That is, it is what would be expected from an isotropic expansion of the cell. However, for the sites that are close to the defect (distance $<$ 5 \AA) the average change in neighboring distances is much larger than 0.6\%. Similarly, for a partially localized hole on U site ($h^{\cdot}$ or U$_{\mathrm{U}}^{\delta+}$) (Fig. \ref{fig:7}(b)), the structure exhibits a large inward relaxation of the lattice close to the defect site, and a constant inward relaxation far away from the defect site, equal to the fractional volume change (-0.6\%) due to removal of an electron from bulk UO$_2$. This is expected in UO$_2$ since charges are known to form small polarons, which by definition carry local lattice distortions\cite{Stoneham2007}.\par

In contrast, for GaAs and Si the electron (or hole) is completely delocalized and does not correspond to any specific atomic site; therefore the average change in the neighboring distance is independent of the site considered: an electron (or a hole) induces a constant expansion in the lattice that is uniform over all atoms. The structure analysis of electron (or hole) in GaAs (Fig. \ref{fig:7}(c) and (d)) shows that the average change in neighboring distances between the defect and bulk structures is a constant value and is equal to the fractional volume change (due to addition of electron or hole). The structure with an extra electron shows a homogeneous expansion of $\sim$0.49\% (Fig. \ref{fig:7}(c)) while that with an extra hole shows a homogeneous contraction of $\sim$0.40\% of the lattice (Fig. \ref{fig:7}(d)), resulting in roughly $\pm$1\% change respectively, in the lattice parameter.\par

These calculations show that the addition or removal of an electron to or from a bulk lattice can result in delocalized or localized charge depending on the system. It is intriguing to observe that independent of the behavior (localized or delocalized) of the charge, there is an expansion (or contraction) of the bulk lattice, resulting in significant relaxation volumes for electrons and holes simulated under DFT framework. Whether such an effect is real or a behavior particular to DFT calculations is not clear at present. However, a detailed examination like this of electrons and holes in bulk structures helps us to establish and understand the connection between relaxation volumes and local structure of charged bulk.\par

\begin{figure*}[t]
\includegraphics[width=0.8\textwidth]{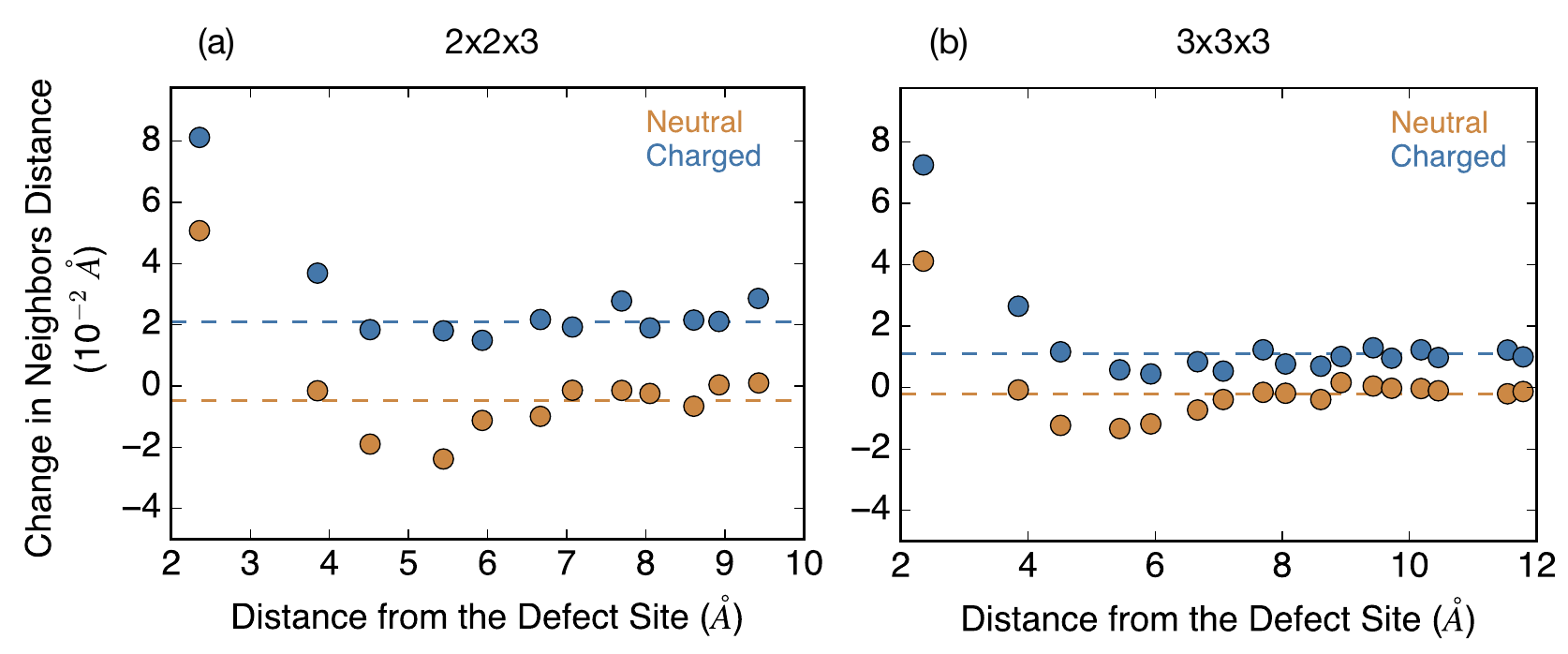}
\caption{\label{fig:8} Change in nearest neighbor distance (solid circles) induced by the uranium vacancy in the neutral (blue) and charged (4-) (yellow) state in: (a) $2\times2\times3$ and (b) $3\times3\times3$ DFT supercell.}
\end{figure*}

\subsubsection{\label{sec:IIIB2}U vacancy in UO$_{2}$}

The U vacancy in the fluorite structure resides at the center of the cube formed by the first eight neighboring oxygen atoms. The difference between neutral and charged supercell calculations lies in the presence or absence of local charge compensation mechanisms around the defect site\cite{Andersson2011, Dorado2012}. We examine these differences by explicitly analyzing the difference in charge density and Bader charge in neutral and charged supercell calculations. The neutral uranium vacancy is charge compensated by four U sites ($h^{\cdot}$ or U$_{\mathrm{U}}^{\delta+}$) with Bader charges of 2.79-2.81 (larger than Bader charge of 2.54 on U site in bulk UO$_2$) located at 2nd (3.86 \AA) and 4th (5.44 \AA) neighboring distances from the defect site. However for the charged (4-) uranium vacancy there is no local charge compensation from neighboring U sites.\par

The changes in nearest neighbor distances (Fig. \ref{fig:8}) for the both the neutral and charged U vacancy clearly reveal the differences in the short and long-range atomic relaxation around the defect site. The lattice distortion around the neutral vacancy extends to a relatively larger distance ($\sim$6-7 \AA) compared to the charged vacancy ($\sim$ 4-5 \AA), because of the presence of neighboring U$_{\mathrm{U}}^{\delta+}$, which are absent for the charged vacancy. For both the neutral and charged supercell, the change in neighboring distances saturates to their respective fractional volume change away from the defect site ($-$0.46\% and 2.1\% for the neutral and charged U vacancy, respectively, in a 144-atom supercell). Analysis in a larger 324-atom supercell yields similar trends in the local and long-range defect structure of neutral and charged U vacancies.\par

In corresponding investigations of charged vacancies in GaAs and Si we find that vacancies show short-ranged local distortion around the localized defect sites. However, changes in nearest neighbor distances away from the localized defect site are manifestations of the overall volume change, which depend on the charge state of the defect.\par

\subsection{\label{sec:IIIC}Defect reactions vs. individual defects}

In the preceding discussion, we have focused on the properties of isolated point defects. However, in reality, defects are never created in isolation, but via a series of reactions involving multiple defects that can be described via Kr\"{o}ger-Vink notation\cite{Kroger1956}. These reactions obey strict conservation rules, namely: mass, charge, and site (maintenance of constant ratio of cationic to anionic sites in the structure) balance. There have been several theoretical studies\cite{Crocombette2001, Andersson2011, Dorado2012} that develop a point defect reaction model for UO$_2$ in order to predict defect concentration, formation energies, and transport properties under thermodynamic conditions prevailing in experiments\cite{Dorado2011}. In this section, we employ a similar approach based on a point defect model to compute defect reaction volumes of the net reaction, as opposed to the individual constituents. We illustrate this by using the example of the oxygen interstitial and the uranium vacancy, which are more likely to form in UO$_2$ under oxidizing/transmutation conditions.\par

The reaction for incorporating oxygen at an interstitial site is given as

\begin{equation}\label{eq:12}
\frac{1}{2}U_{3}O_{8} (\mathrm{s}) + V_{\mathrm{I}}^{\times} \rightleftharpoons O_{\mathrm{I}}^{''} + 2h^{\cdot} +  \frac{3}{2}UO_{2} (\mathrm{s}) ,
\end{equation}
where $V_{\mathrm{I}}^{\times}$ is a vacant octahedral interstitial site, in the fluorite structure, $O_{\mathrm{I}}^{''}$ is the oxygen ion at the interstitial site, and $h^{\cdot}$ represents an U$^{5+}$ ion (hole) not bounded to the oxygen interstitial $O_{\mathrm{I}}^{''}$. The volume change, $\Delta \Omega^{\mathrm{rxn}}$, for the reaction (Eq. \ref{eq:12}) is then given as the sum of the individual relaxation volumes, $\Delta \Omega^{r}$, obtained from separate DFT supercell calculations of the charged (2-) oxygen interstitial and the U$^{5+}$ ion,

\begin{equation}\label{eq:13}
\Delta \Omega_{O_{\mathrm{I}}}^{\mathrm{rxn}} = \Delta \Omega^{r} (O_{\mathrm{I}}^{''}) + 2\Delta \Omega^{r}(h^{\cdot})
+ \frac{3}{2}\Omega_{UO_{2}} - \frac{1}{2}\Omega_{U_{3}O_{8}} ,
\end{equation}
The reaction volume $\Delta \Omega_{O_{\mathrm{I}}}^{\mathrm{rxn}} $ for the oxygen interstitial reaction is computed to be $-$25.8 \AA$^{3}$, implying contraction of the lattice due to formation of this overall neutral collection of defects (note that, in this calculation, the interstitial and the holes are assumed to be separated to infinity). The sum of the relaxation volumes of charged oxygen interstitial and holes is $-$3.2 \AA$^{3}$, similar to the relaxation volume, $\Delta \Omega^{r} (O_{\mathrm{I}}^{\times})$ of a neutral oxygen interstitial of $-$3.38 \AA$^{3}$ calculated within a DFT supercell, which is not surprising if neutral O interstitial can be understood as charged oxygen interstitial bounded by two holes.\par

Similarly, a defect reaction for the formation of a uranium vacancy, is given as

\begin{equation}\label{eq:14}
U_{3}O_{8} (\mathrm{s}) + U_{\mathrm{U}}^{\times} \rightleftharpoons V_{\mathrm{U}}^{''''} + 4h^{\cdot} +  4(UO_{2})_{\mathrm{SRG}} ,
\end{equation}
where point defects $V_{\mathrm{U}}^{''''}$ and $h^{\cdot}$ are the fully charged uranium vacancy and U$^{5+}$ ion (hole), respectively, and (UO$_2$)$_{\mathrm{SRG}}$ represents the corresponding lattice atoms present at a site of repeatable growth (SRG) such as a surface, dislocation, grain boundary. The volume change for the reaction (Eq. \ref{eq:14}) is then given as

\begin{equation}\label{eq:15}
\Delta \Omega_{V_{\mathrm{U}}}^{\mathrm{rxn}} = \Delta \Omega^{r} (V_{\mathrm{U}}^{''''}) + 4\Delta \Omega^{r}(h^{\cdot})
+ 4\Omega_{UO_{2}} - \Omega_{U_{3}O_{8}} ,
\end{equation}
involving individual charged-defect relaxation volumes $\Delta \Omega^{r} (V_{\mathrm{U}}^{''''})$ and $\Delta \Omega^{r}(h^{\cdot})$, calculated again from separate supercell calculations. The overall computed reaction volume $\Delta \Omega_{V_{\mathrm{U}}}^{\mathrm{rxn}}$ is $-$10.8 \AA$^{3}$, which again suggests contraction of the lattice due to the formation of uranium vacancies. These findings are in good agreement with the experimental results\cite{} and with theoretical prediction\cite{} for the contraction of the UO$_{2}$ lattice due to the formation of the hyper-stoichiometric phase UO$_{2+x}$ (with x $\textless$ 0.5). The sum of the relaxation volumes of individual charge defects (first two terms in Eq. \ref{eq:15}) is about $-$6.26 \AA$^{3}$, similar but not exactly equal to that predicted by the relaxation of neutral U vacancy ($\Delta \Omega^{r} (V_{\mathrm{U}}^{\times})$ = $-$9.07 \AA$^{3}$) in a DFT supercell calculation.\par

Our results comparing the relaxation volume of the neutral defects with the summed relaxation volumes of the charged defects calculated from separate supercells calculations are shown in Table \ref{tab:6}. There is good agreement with regards to the sign of the lattice relaxation (expansion or contraction) due to the formation of an overall neutral defect reaction with the two approaches. The difference in the predicted magnitude can be attributed to the interaction (both elastic and electrostatic) between the individual charged defects. Thus, while the relaxation volumes of the charged defects have large magnitudes, they cancel when summed, leading to values that are in reasonable agreement with corresponding neutral defects. Similarly, while defect reactions, such as those in Eqs. \ref{eq:12} and \ref{eq:14}, contain components corresponding to charged defects, once these are summed in a net neutral reaction, the large volumes of the charged defects cancel, leading to relatively small volumes for the reaction itself.\par

\begin{table}[t]
	\caption{\label{tab:6} Comparison of the defect relaxation volume (\AA$^{3}$) of neutral individual defects computed within a single (one) supercell (labeled OS) or via a sum of defect relaxation volumes of individual charge defects computed in separate supercells.}
		\begin{ruledtabular}
		\begin{tabular}{ldld}
		Defect &  \multicolumn{1}{c}{$\sum \Delta \Omega^{r}$} & Defect & \multicolumn{1}{c}{$\Delta \Omega^{r}$} \\[0.2em]
		\colrule \\ [-0.7em]
		$V_{\mathrm{O}}^{\cdot \cdot} + O_{\mathrm{I}}^{''}$	&3.26		&$V_{\mathrm{O}}^{\cdot \cdot} + O_{\mathrm{I}}^{''}$ (OS)		&3.63	\\ [0.3em]
		$V_{\mathrm{U}}^{''''} + 4h^{\cdot}$				&-6.26		&$V_{\mathrm{U}}^{\times}$								&-9.07	\\ [0.3em]
		$O_{\mathrm{I}}^{''} + 2h^{\cdot}$				&-3.20		&$O_{\mathrm{I}}^{\times}$								&-3.38	\\ [0.3em]
		$V_{\mathrm{O}}^{\cdot \cdot} + 2e^{''}$			&6.52		&$V_{\mathrm{O}}^{\times}$								&0.56	\\ [0.3em]
		$V_{\mathrm{Ga}}^{'''} + 3h^{\cdot}$				&-33.83		&$V_{\mathrm{Ga}}^{\times}$								&-26.29	\\ [0.3em]
		$V_{\mathrm{Si}}^{\cdot \cdot} + 2e^{'}$			&-12.39		&$V_{\mathrm{Si}}^{\times}$								&-14.69	\\
		\end{tabular}
	\end{ruledtabular}
\end{table}
			
\section{\label{sec:IV} Discussion}

In calculating the elastic dipole tensor for individual charged defects, we have found that the magnitude of these dipole tensors lead to seemingly unphysical behavior. For example, as noted in the results, segregation energies of defects to dislocations are extremely large when considering the properties of individual charged defects. In an effort to understand the origin of this behavior in charged point defects, we have applied volume correction schemes that are analogous to those proposed for correcting the energetics of charged defects. While these corrections can have large effects on the energetics of defects, our results indicate that corresponding changes to defect volumes (and thus to dipole tensors) are relatively small and are not the source of the large dipole tensors and the defect relaxation volumes.\par

\begin{figure*}[t]
\includegraphics[width=0.9\textwidth]{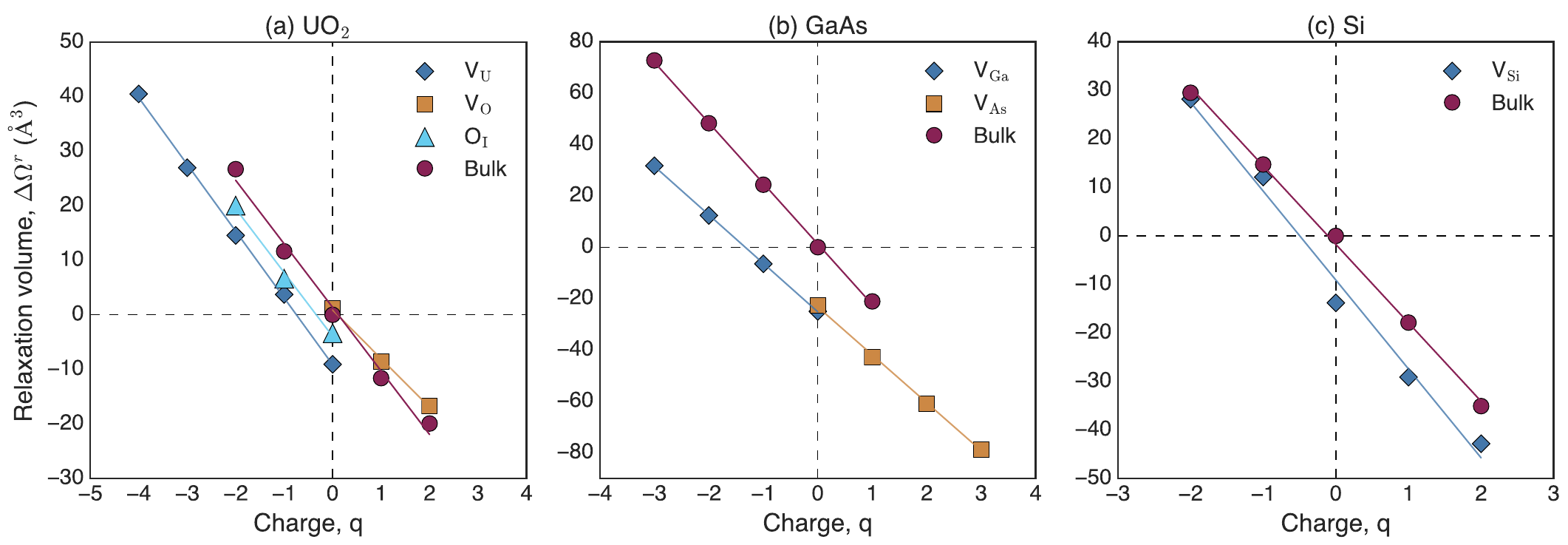}
\caption{\label{fig:9} Defect relaxation volume as a function of the overall charge on the DFT supercell for point defects and bulk in (a) UO2, (b) GaAs, and (c) Si.}
\end{figure*}

In contrast, if one considers a full defect reaction and the associated relaxation volume, while the individual contributions correspond to charged defects and thus have large relaxation volumes, the volume change of the overall reactions are physically reasonable and are similar to those of neutral defects. Thus, the large relaxation volumes of negative defects are essentially canceled out by large, but oppositely signed, relaxation volumes of positive defects. This suggests that one cannot think of a charged defect in isolation but only in the context of the overall defect reactions that produce the defect in the first place. However, in numerous fields, we are accustomed to thinking of the properties of isolated charged defects as entities that migrate and interact in the material. For example, in fast ion conductors, we consider the transport of charged carriers in the material as specific entities. Thus, it seems reasonable that one should be able to describe the elastic response of the medium due to a charged defect and thus determine how it interacts with, for example, dislocations.\par

Therefore, the above discussion suggests that, perhaps, the problem lies with the reference state reservoir. The defect formation volume depends on the choice of reservoir, just as does the defect formation energy. Considering a neutral defect, the atoms removed or added from the perfect bulk material have to be placed in a reservoir and the volume of both the material and the reservoir changes. Analogously, for charged defects, the electrons or holes added to the defect system also contribute to both the defect formation energy and volume. While the contribution to the defect formation energy has been well studied and several correction terms have been developed, the contribution of the change in charge to the volume of the reservoir has not been considered.\par

\begin{table*}[t]
	\caption{\label{tab:7} Comparison of the corrected defect formation volumes $\Delta \Omega ^{f}$ (\AA$^{3}$) of neutral and charged individual defects, with the defect relaxation volumes $\Delta \Omega ^{\mathrm{rxn}}$.}
		\begin{ruledtabular}
		\begin{tabular}{cccccc}
		\\[-0.7em]
		Defect Reaction	&  \multicolumn{1}{c}{$\Delta \Omega^{\mathrm{rxn}}$} & Individual Charged Defect & \multicolumn{1}{c}{$\Delta \Omega^{f}$} & Individual Neutral Defect & \multicolumn{1}{c}{$\Delta \Omega^{f}$}
		\\[0.2em]
		\colrule \\ [-0.7em]
		$V_{\mathrm{U}}^{''''}$ (Eq.\ref{eq:14})	&-10.84	&$V_{\mathrm{U}}^{''''}$		&-10.80		&$V_{\mathrm{U}}^{\times}$		&-13.97	\\ [0.3em]
		$O_{\mathrm{I}}^{''}$	 (Eq.\ref{eq:12})	&-25.80	&$O_{\mathrm{I}}^{''}$		&-25.83		&$O_{\mathrm{I}}^{\times}$		&-26.01	\\ [0.3em]
		$V_{\mathrm{O}}^{\cdot \cdot}$		&29.15	&$V_{\mathrm{O}}^{\cdot \cdot}$	&29.13		&$V_{\mathrm{O}}^{\times}$		&23.19	\\ [0.3em]
		$V_{\mathrm{Ga}}^{'''}$			&-13.49	&$V_{\mathrm{Ga}}^{'''}$			&-16.26		&$V_{\mathrm{Ga}}^{\times}$		&-5.95	\\ [0.3em]
		$V_{\mathrm{Si}}^{\cdot \cdot}$		&8.05	&$V_{\mathrm{Si}}^{\cdot \cdot}$	&9.92		&$V_{\mathrm{Si}}^{\times}$		&5.75	\\
		\end{tabular}
	\end{ruledtabular}
\end{table*}

\begin{table*}[t]
	\caption{\label{tab:8s} Comparison of the corrected defect relaxation volume $\big[ \Delta \Omega^{r} - q\Delta \Omega^{\mathrm{ref}} \big]$ (\AA$^{3}$) of neutral and charged individual defects, with the sum of defect relaxation volumes of individual charge defects computed in separate supercells under the defect reaction approach. (Some of these values appeared in Table \ref{tab:6} but are reproduced here for convenience.)}
		\begin{ruledtabular}
		\begin{tabular}{cccccc}
		\\ [-0.7em]
		Defect Reaction	&  \multicolumn{1}{c}{$\sum \Delta \Omega^{r}$} & Individual Charged Defect & \multicolumn{1}{c}{$\Delta \Omega^{r} -q\Delta \Omega^{\mathrm{ref}}$} & Individual Neutral Defect & \multicolumn{1}{c}{$\Delta \Omega^{r} -q\Delta \Omega^{\mathrm{ref}}$}\\[0.2em]
		\colrule \\ [-0.7em]
		$V_{\mathrm{U}}^{''''} + 4h^{\cdot}$		&-6.26	&$V_{\mathrm{U}}^{''''}$	&-6.02	&$V_{\mathrm{U}}^{\times}$	&-9.07	\\ [0.3em]
		$O_{\mathrm{I}}^{''} + 2h^{\cdot}$		&-3.20	&$O_{\mathrm{I}}^{''}$	&-3.28	&$O_{\mathrm{I}}^{\times}$	&-3.38	\\ [0.3em]
		$V_{\mathrm{O}}^{\cdot \cdot} + 2e^{''}$	&6.52	&$V_{\mathrm{O}}^{\cdot \cdot}$	&6.54	&$V_{\mathrm{O}}^{\times}$	&0.56	\\ [0.3em]
		$V_{\mathrm{Ga}}^{'''} + 3h^{\cdot}$		&-33.83	&$V_{\mathrm{Ga}}^{'''}$	&-36.07	&$V_{\mathrm{Ga}}^{\times}$	&-26.29	\\ [0.3em]
		$V_{\mathrm{Si}}^{\cdot \cdot} + 2e^{'}$	&-12.39	&$V_{\mathrm{Si}}^{\cdot \cdot}$	&-10.57	&$V_{\mathrm{Si}}^{\times}$	&-14.69	\\
		\end{tabular}
	\end{ruledtabular}
\end{table*}

From our results for three different materials systems, we can estimate the contribution of the net charge to the defect formation volume, as shown in Fig. \ref{fig:9}. We see a linear relationship between the electron count and the defect relaxation volume in all three systems, similar to that observed in previous DFT studies\cite{Centoni2005, Windl2002, Bruneval2012}. Defect formation volumes, as defined in Eq. \ref{eq:5}, require $\Omega_{0}$ to be added or subtracted to the defect relaxation volume to account for the volume of the individual species added or removed from the reservoir to maintain overall mass balance. However, in Eq. \ref{eq:5}, the contribution of the change in charge to the volume of the electron reservoir is missing. The change in the volume of the bulk with charge can be attributed to the volume contribution per electron as coming from the change in volume of the electron reservoir, $\Delta \Omega^{\mathrm{ref}}$. Therefore, a complete definition of defect formation volume should be given as

\begin{equation}\label{eq:16}
\Delta \Omega^{f} (X,q) = \Delta \Omega^{r} (X,q) -q\Delta \Omega^{\mathrm{ref}} \pm \Omega_{0} .
\end{equation}
$\Delta \Omega^{\mathrm{ref}}$ is determined by the slope of the bulk relaxation volume as a function of charge. Value of $\Delta \Omega^{\mathrm{ref}}$ is system dependent and for UO$_2$, GaAs, and Si we find $-$11.6 \AA$^{3}$, $-$23.6 $^{3}$ and $-$16.14 \AA$^{3}$ per unit charge, respectively. For Si, Windl and Daw\cite{Windl2002} reported a value of about $-$15 \AA$^{3}$ from the slope of the change in volume of the bulk with net charge, similar to the value we have obtained. One can then recompute defect formation volumes by accounting for $\Delta \Omega^{\mathrm{ref}}$ in all three systems, as summarized in Table \ref{tab:7}. Reaction volumes are not affected by $\Delta \Omega^{\mathrm{ref}}$ as the reaction is overall charge neutral and, hence, the change in volume of the electronic reservoir due to charge cancels out exactly.\par

The charged defect elastic dipole tensor is theoretically computed from its relaxation volume (Eq. \ref{eq:1}), and experimental work often report it as chemical strain\cite{Freedman2009, Nazarov2016}, estimated by measuring the change in the lattice parameter as function of deviation from the stoichiometric chemical formula. The apparent question that comes next is should the charged defect relaxation volume be corrected to account for the volume associated with the electron/hole reservoir.\par

The change in the volume of the bulk with charge (volume associated with the electron/hole reservoir) as given by $\Delta \Omega^{\mathrm{ref}}$ can be understood as the lattice perturbation associated with excess electron and/or holes in the bulk, as described in Sec. \ref{sec:IIIB1}. We reason that charged defects are individual entities that are present in an overall charge neutral system. Both their elastic dipole tensor and relaxation volumes are meaningful when defined with respect to the electron/hole reservoir in order to keep the overall system charge neutral, as one would expect when measuring these defect properties from experiments. Traditionally, when one calculates the elastic dipole tensor and relaxation volume of a charged defect, it is relative to the neutral bulk. Perhaps the correct reference for a charged atomic defect in a lattice is not the neutral bulk, but the charged bulk, so that the defect represents a perturbation of the charged lattice only, and not the neutral lattice.\par

The elastic dipole tensor $G_{ij}$ for a charged defect when defined in an overall neutral (with respect to the electron or hole reservoir) but locally charged lattice can then be given as 

\begin{equation}\label{eq:17}
G_{ij} = C_{ijkl} \Big[\Delta\Omega_{kl}^{r} - q\Delta \Omega^{\mathrm{ref}} \big] ,
\end{equation}
where, $[\Delta\Omega_{kl}^{r} - q\Delta \Omega^{\mathrm{ref}} \big]$  represents the corrected relaxation volume. As listed in Table \ref{tab:8}, the corrected relaxation volume of charged defects are more reasonable in magnitude when compared to relaxation volumes for both the neutral individual defects and those of defect reactions. For example, the corrected relaxation volume of $-$3.28 \AA$^3$ for the charged (2$-$) oxygen interstitial (overall neutral, i.e. the oxygen interstitial with respect to doubly charged bulk UO$_{2}$) compares very well with the volume of $-$3.20 \AA$^3$ of the reaction involving charged (2$-$) oxygen interstitial with two holes. Similar comparisons for other point defects in UO$_2$ are also very good, because as can be seen in Fig. \ref{fig:9}, the slopes of change in volume of defects with charge are almost parallel to the slope of change in volume of bulk with charge. However, for Si and GaAs we observe that different defects have different slopes, suggesting that the effect of electrons and holes present in different defects leads to different strains and volume changes.\par

Finally, computing the dipole tensor (Eq. \ref{eq:17}) of charged defects from the corrected relaxation volumes will result in physically reasonable magnitude in the change in defect formation energy of the charged point defect upon interaction with external strains. Accounting for the change in volume of the electron reservoir with net charge in the definition of the defect formation volume and elastic dipole tensor allows us to reconcile these two perspectives: that defect behavior not only has meaning via reactions but that isolated defects in an overall neutral lattice (which requires the reservoir) also have unique properties that determine how they behave within the lattice. 

\section{\label{sec:V} Conclusions}

We have shown that point defect relaxation volumes and the associated defect elastic dipole tensors, which measure the elastic distortion of the bulk lattice caused by the presence of the defect, are peculiarly large for charged defects. Defect relaxation volumes have a finite measure even for infinitely large system sizes, and we show that the large relaxation volume of charged defects is not due to finite-size artifacts within the DFT supercell approach, as previously suspected. Relaxation volumes using an overall neutral defect reaction model based on DFT calculations of isolated charged defects agree reasonably well in terms of the overall nature of lattice relaxation and magnitude when compared with the individual neutral defects simulated within a single DFT supercell. We postulate that, similar to the defect formation energies, the defect formation volumes also depend on the choice of reservoir. We show that by taking into account the contribution of change in charge to the volume of the electron reservoir, the recomputed charge defect formation volumes and elastic dipole tensors are reasonable in magnitude, and that it is possible to describe the elastic properties of isolated charge defects within the material, not only in the context of the overall defect reactions that produce the defect but also as individual species in an overall neutral lattice.

\begin{acknowledgments}
AG is happy to acknowledge the Materials Science and Technology Division at Los Alamos National Laboratory for hospitality during this project. We also like to thank Georg Kresse and Christoph Freysoldt for their guidance and useful discussion on the charged defect correction schemes. The work of AG and SRP was supported by the United States Department of Energy, Office of Nuclear Energy (DOE-NE) Nuclear Energy University Program 10-2258 and the work of RGH was supported by the NSF CAREER award DMR-1542776.  Work at LANL was funded by the DOE Nuclear Energy Advanced Modeling and Simulation (NEAMS) program.
\end{acknowledgments}

\bibliography{manuscript}

\begin{thebibliography}{66}%
\makeatletter
\providecommand \@ifxundefined [1]{%
 \@ifx{#1\undefined}
}%
\providecommand \@ifnum [1]{%
 \ifnum #1\expandafter \@firstoftwo
 \else \expandafter \@secondoftwo
 \fi
}%
\providecommand \@ifx [1]{%
 \ifx #1\expandafter \@firstoftwo
 \else \expandafter \@secondoftwo
 \fi
}%
\providecommand \natexlab [1]{#1}%
\providecommand \enquote  [1]{``#1''}%
\providecommand \bibnamefont  [1]{#1}%
\providecommand \bibfnamefont [1]{#1}%
\providecommand \citenamefont [1]{#1}%
\providecommand \href@noop [0]{\@secondoftwo}%
\providecommand \href [0]{\begingroup \@sanitize@url \@href}%
\providecommand \@href[1]{\@@startlink{#1}\@@href}%
\providecommand \@@href[1]{\endgroup#1\@@endlink}%
\providecommand \@sanitize@url [0]{\catcode `\\12\catcode `\$12\catcode
  `\&12\catcode `\#12\catcode `\^12\catcode `\_12\catcode `\%12\relax}%
\providecommand \@@startlink[1]{}%
\providecommand \@@endlink[0]{}%
\providecommand \url  [0]{\begingroup\@sanitize@url \@url }%
\providecommand \@url [1]{\endgroup\@href {#1}{\urlprefix }}%
\providecommand \urlprefix  [0]{URL }%
\providecommand \Eprint [0]{\href }%
\providecommand \doibase [0]{http://dx.doi.org/}%
\providecommand \selectlanguage [0]{\@gobble}%
\providecommand \bibinfo  [0]{\@secondoftwo}%
\providecommand \bibfield  [0]{\@secondoftwo}%
\providecommand \translation [1]{[#1]}%
\providecommand \BibitemOpen [0]{}%
\providecommand \bibitemStop [0]{}%
\providecommand \bibitemNoStop [0]{.\EOS\space}%
\providecommand \EOS [0]{\spacefactor3000\relax}%
\providecommand \BibitemShut  [1]{\csname bibitem#1\endcsname}%
\let\auto@bib@innerbib\@empty
\bibitem [{\citenamefont {Seebauer}\ and\ \citenamefont
  {Kratzer}(2006)}]{Seebauer2006}%
  \BibitemOpen
  \bibfield  {author} {\bibinfo {author} {\bibfnamefont {E.~G.}\ \bibnamefont
  {Seebauer}}\ and\ \bibinfo {author} {\bibfnamefont {M.~C.}\ \bibnamefont
  {Kratzer}},\ }\href {\doibase 10.1016/j.mser.2006.01.002} {\bibfield
  {journal} {\bibinfo  {journal} {Materials Science and Engineering R:
  Reports}\ }\textbf {\bibinfo {volume} {55}},\ \bibinfo {pages} {57} (\bibinfo
  {year} {2006})}\BibitemShut {NoStop}%
\bibitem [{\citenamefont {Was}(2007)}]{Was2007}%
  \BibitemOpen
  \bibfield  {author} {\bibinfo {author} {\bibfnamefont {G.~S.}\ \bibnamefont
  {Was}},\ }\href {http://books.google.com/books?id=0N06swAJI0AC} {\emph
  {\bibinfo {title} {{Fundamentals of Radiation Materials Science}}}}\
  (\bibinfo  {publisher} {Springer},\ \bibinfo {year} {2007})\BibitemShut
  {NoStop}%
\bibitem [{\citenamefont {Gillan}(1981)}]{Gillan1981}%
  \BibitemOpen
  \bibfield  {author} {\bibinfo {author} {\bibfnamefont {M.~J.}\ \bibnamefont
  {Gillan}},\ }\href {\doibase 10.1080/01418618108239410} {\bibfield  {journal}
  {\bibinfo  {journal} {Philosophical Magazine A}\ }\textbf {\bibinfo {volume}
  {43}},\ \bibinfo {pages} {301} (\bibinfo {year} {1981})}\BibitemShut
  {NoStop}%
\bibitem [{\citenamefont {Lidiard}(1981)}]{Lidiard1981}%
  \BibitemOpen
  \bibfield  {author} {\bibinfo {author} {\bibfnamefont {A.}~\bibnamefont
  {Lidiard}},\ }\href
  {http://www.tandfonline.com/doi/abs/10.1080/01418618108239409} {\bibfield
  {journal} {\bibinfo  {journal} {Philosophical Magazine A}\ }\textbf {\bibinfo
  {volume} {43}},\ \bibinfo {pages} {291} (\bibinfo {year} {1981})}\BibitemShut
  {NoStop}%
\bibitem [{\citenamefont {Nazarov}\ \emph {et~al.}(2016)\citenamefont
  {Nazarov}, \citenamefont {Majevadia}, \citenamefont {Patel}, \citenamefont
  {Wenman}, \citenamefont {Balint}, \citenamefont {Neugebauer},\ and\
  \citenamefont {Sutton}}]{Nazarov2016}%
  \BibitemOpen
  \bibfield  {author} {\bibinfo {author} {\bibfnamefont {R.}~\bibnamefont
  {Nazarov}}, \bibinfo {author} {\bibfnamefont {J.~S.}\ \bibnamefont
  {Majevadia}}, \bibinfo {author} {\bibfnamefont {M.}~\bibnamefont {Patel}},
  \bibinfo {author} {\bibfnamefont {M.~R.}\ \bibnamefont {Wenman}}, \bibinfo
  {author} {\bibfnamefont {D.~S.}\ \bibnamefont {Balint}}, \bibinfo {author}
  {\bibfnamefont {J.}~\bibnamefont {Neugebauer}}, \ and\ \bibinfo {author}
  {\bibfnamefont {A.~P.}\ \bibnamefont {Sutton}},\ }\href {\doibase
  10.1103/PhysRevB.94.241112} {\bibfield  {journal} {\bibinfo  {journal}
  {Physical Review B}\ }\textbf {\bibinfo {volume} {94}},\ \bibinfo {pages}
  {241112} (\bibinfo {year} {2016})}\BibitemShut {NoStop}%
\bibitem [{\citenamefont {Clouet}\ \emph {et~al.}(2008)\citenamefont {Clouet},
  \citenamefont {Garruchet}, \citenamefont {Nguyen}, \citenamefont {Perez},\
  and\ \citenamefont {Becquart}}]{Clouet2008}%
  \BibitemOpen
  \bibfield  {author} {\bibinfo {author} {\bibfnamefont {E.}~\bibnamefont
  {Clouet}}, \bibinfo {author} {\bibfnamefont {S.}~\bibnamefont {Garruchet}},
  \bibinfo {author} {\bibfnamefont {H.}~\bibnamefont {Nguyen}}, \bibinfo
  {author} {\bibfnamefont {M.}~\bibnamefont {Perez}}, \ and\ \bibinfo {author}
  {\bibfnamefont {C.~S.}\ \bibnamefont {Becquart}},\ }\href {\doibase
  10.1016/j.actamat.2008.03.024} {\bibfield  {journal} {\bibinfo  {journal}
  {Acta Materialia}\ }\textbf {\bibinfo {volume} {56}},\ \bibinfo {pages}
  {3450} (\bibinfo {year} {2008})}\BibitemShut {NoStop}%
\bibitem [{\citenamefont {Goyal}\ \emph {et~al.}(2015)\citenamefont {Goyal},
  \citenamefont {Phillpot}, \citenamefont {Subramanian}, \citenamefont
  {Andersson}, \citenamefont {Stanek},\ and\ \citenamefont
  {Uberuaga}}]{Goyal2015}%
  \BibitemOpen
  \bibfield  {author} {\bibinfo {author} {\bibfnamefont {A.}~\bibnamefont
  {Goyal}}, \bibinfo {author} {\bibfnamefont {S.~R.}\ \bibnamefont {Phillpot}},
  \bibinfo {author} {\bibfnamefont {G.}~\bibnamefont {Subramanian}}, \bibinfo
  {author} {\bibfnamefont {D.~A.}\ \bibnamefont {Andersson}}, \bibinfo {author}
  {\bibfnamefont {C.~R.}\ \bibnamefont {Stanek}}, \ and\ \bibinfo {author}
  {\bibfnamefont {B.~P.}\ \bibnamefont {Uberuaga}},\ }\href {\doibase
  10.1103/PhysRevB.91.094103} {\bibfield  {journal} {\bibinfo  {journal}
  {Physical Review B}\ }\textbf {\bibinfo {volume} {91}},\ \bibinfo {pages}
  {094103} (\bibinfo {year} {2015})}\BibitemShut {NoStop}%
\bibitem [{\citenamefont {Centoni}\ \emph {et~al.}(2005)\citenamefont
  {Centoni}, \citenamefont {Sadigh}, \citenamefont {Gilmer}, \citenamefont
  {Lenosky}, \citenamefont {{D{\'{i}}az de la Rubia}},\ and\ \citenamefont
  {Musgrave}}]{Centoni2005}%
  \BibitemOpen
  \bibfield  {author} {\bibinfo {author} {\bibfnamefont {S.}~\bibnamefont
  {Centoni}}, \bibinfo {author} {\bibfnamefont {B.}~\bibnamefont {Sadigh}},
  \bibinfo {author} {\bibfnamefont {G.}~\bibnamefont {Gilmer}}, \bibinfo
  {author} {\bibfnamefont {T.}~\bibnamefont {Lenosky}}, \bibinfo {author}
  {\bibfnamefont {T.}~\bibnamefont {{D{\'{i}}az de la Rubia}}}, \ and\ \bibinfo
  {author} {\bibfnamefont {C.}~\bibnamefont {Musgrave}},\ }\href {\doibase
  10.1103/PhysRevB.72.195206} {\bibfield  {journal} {\bibinfo  {journal}
  {Physical Review B}\ }\textbf {\bibinfo {volume} {72}},\ \bibinfo {pages}
  {195206} (\bibinfo {year} {2005})}\BibitemShut {NoStop}%
\bibitem [{\citenamefont {Windl}\ and\ \citenamefont
  {Daw}(2002{\natexlab{a}})}]{Windl2002a}%
  \BibitemOpen
  \bibfield  {author} {\bibinfo {author} {\bibfnamefont {W.}~\bibnamefont
  {Windl}}\ and\ \bibinfo {author} {\bibfnamefont {M.}~\bibnamefont {Daw}},\
  }\href@noop {} {\bibfield  {journal} {\bibinfo  {journal} {Nanotech}\
  }\textbf {\bibinfo {volume} {2}},\ \bibinfo {pages} {197} (\bibinfo {year}
  {2002}{\natexlab{a}})}\BibitemShut {NoStop}%
\bibitem [{\citenamefont {Bruneval}\ and\ \citenamefont
  {Crocombette}(2012)}]{Bruneval2012}%
  \BibitemOpen
  \bibfield  {author} {\bibinfo {author} {\bibfnamefont {F.}~\bibnamefont
  {Bruneval}}\ and\ \bibinfo {author} {\bibfnamefont {J.-P.}\ \bibnamefont
  {Crocombette}},\ }\href {\doibase 10.1103/PhysRevB.86.140103} {\bibfield
  {journal} {\bibinfo  {journal} {Physical Review B}\ }\textbf {\bibinfo
  {volume} {86}},\ \bibinfo {pages} {140103} (\bibinfo {year}
  {2012})}\BibitemShut {NoStop}%
\bibitem [{\citenamefont {Bruneval}\ \emph {et~al.}(2015)\citenamefont
  {Bruneval}, \citenamefont {Varvenne}, \citenamefont {Crocombette},\ and\
  \citenamefont {Clouet}}]{Bruneval2015}%
  \BibitemOpen
  \bibfield  {author} {\bibinfo {author} {\bibfnamefont {F.}~\bibnamefont
  {Bruneval}}, \bibinfo {author} {\bibfnamefont {C.}~\bibnamefont {Varvenne}},
  \bibinfo {author} {\bibfnamefont {J.-P.}\ \bibnamefont {Crocombette}}, \ and\
  \bibinfo {author} {\bibfnamefont {E.}~\bibnamefont {Clouet}},\ }\href
  {\doibase 10.1103/PhysRevB.91.024107} {\bibfield  {journal} {\bibinfo
  {journal} {Physical Review B}\ }\textbf {\bibinfo {volume} {91}},\ \bibinfo
  {pages} {1} (\bibinfo {year} {2015})}\BibitemShut {NoStop}%
\bibitem [{\citenamefont {Erhart}\ \emph {et~al.}(2006)\citenamefont {Erhart},
  \citenamefont {Albe},\ and\ \citenamefont {Klein}}]{Erhart2006}%
  \BibitemOpen
  \bibfield  {author} {\bibinfo {author} {\bibfnamefont {P.}~\bibnamefont
  {Erhart}}, \bibinfo {author} {\bibfnamefont {K.}~\bibnamefont {Albe}}, \ and\
  \bibinfo {author} {\bibfnamefont {A.}~\bibnamefont {Klein}},\ }\href
  {\doibase 10.1103/PhysRevB.73.205203} {\bibfield  {journal} {\bibinfo
  {journal} {Physical Review B}\ }\textbf {\bibinfo {volume} {73}},\ \bibinfo
  {pages} {205203} (\bibinfo {year} {2006})}\BibitemShut {NoStop}%
\bibitem [{\citenamefont {{\'{A}}goston}\ and\ \citenamefont
  {Albe}(2009)}]{Agoston2009}%
  \BibitemOpen
  \bibfield  {author} {\bibinfo {author} {\bibfnamefont {P.}~\bibnamefont
  {{\'{A}}goston}}\ and\ \bibinfo {author} {\bibfnamefont {K.}~\bibnamefont
  {Albe}},\ }\href {\doibase 10.1039/b900280d} {\bibfield  {journal} {\bibinfo
  {journal} {Physical Chemistry Chemical Physics}\ }\textbf {\bibinfo {volume}
  {11}},\ \bibinfo {pages} {3226} (\bibinfo {year} {2009})}\BibitemShut
  {NoStop}%
\bibitem [{\citenamefont {Grieshammer}\ \emph {et~al.}(2013)\citenamefont
  {Grieshammer}, \citenamefont {Zacherle},\ and\ \citenamefont
  {Martin}}]{Grieshammer2013}%
  \BibitemOpen
  \bibfield  {author} {\bibinfo {author} {\bibfnamefont {S.}~\bibnamefont
  {Grieshammer}}, \bibinfo {author} {\bibfnamefont {T.}~\bibnamefont
  {Zacherle}}, \ and\ \bibinfo {author} {\bibfnamefont {M.}~\bibnamefont
  {Martin}},\ }\href {\doibase 10.1039/c3cp51913a} {\bibfield  {journal}
  {\bibinfo  {journal} {Physical chemistry chemical physics : PCCP}\ }\textbf
  {\bibinfo {volume} {15}},\ \bibinfo {pages} {15935} (\bibinfo {year}
  {2013})}\BibitemShut {NoStop}%
\bibitem [{\citenamefont {Gillan}(1984)}]{Gillan1984}%
  \BibitemOpen
  \bibfield  {author} {\bibinfo {author} {\bibfnamefont {M.~J.}\ \bibnamefont
  {Gillan}},\ }\href {http://iopscience.iop.org/0022-3719/17/9/006} {\bibfield
  {journal} {\bibinfo  {journal} {Journal of Physics C: Solid State Physics}\
  }\textbf {\bibinfo {volume} {17}} (\bibinfo {year} {1984})}\BibitemShut
  {NoStop}%
\bibitem [{\citenamefont {Nowick}\ and\ \citenamefont
  {Berry}(1972)}]{Nowick1972}%
  \BibitemOpen
  \bibfield  {author} {\bibinfo {author} {\bibfnamefont {A.}~\bibnamefont
  {Nowick}}\ and\ \bibinfo {author} {\bibfnamefont {B.}~\bibnamefont {Berry}},\
  }\href
  {http://books.google.com/books?hl=en{\&}lr={\&}id=wO2eVGpROnQC{\&}oi=fnd{\&}pg=PP1{\&}dq=Anelastic+Relaxation+in+Crystalling+Solids{\&}ots=UZlJY8vXxs{\&}sig=um9s6OQFjHl0wUpjnffZY5wO0ao}
  {\emph {\bibinfo {title} {New York, Academic Press}}}\ (\bibinfo  {publisher}
  {Academic Press},\ \bibinfo {address} {New York},\ \bibinfo {year}
  {1972})\BibitemShut {NoStop}%
\bibitem [{\citenamefont {Freedman}\ \emph {et~al.}(2009)\citenamefont
  {Freedman}, \citenamefont {Roundy},\ and\ \citenamefont
  {Arias}}]{Freedman2009}%
  \BibitemOpen
  \bibfield  {author} {\bibinfo {author} {\bibfnamefont {D.}~\bibnamefont
  {Freedman}}, \bibinfo {author} {\bibfnamefont {D.}~\bibnamefont {Roundy}}, \
  and\ \bibinfo {author} {\bibfnamefont {T.}~\bibnamefont {Arias}},\ }\href
  {\doibase 10.1103/PhysRevB.80.064108} {\bibfield  {journal} {\bibinfo
  {journal} {Physical Review B}\ }\textbf {\bibinfo {volume} {80}},\ \bibinfo
  {pages} {064108} (\bibinfo {year} {2009})}\BibitemShut {NoStop}%
\bibitem [{\citenamefont {Puchala}\ \emph {et~al.}(2008)\citenamefont
  {Puchala}, \citenamefont {Falk},\ and\ \citenamefont
  {Garikipati}}]{Puchala2008}%
  \BibitemOpen
  \bibfield  {author} {\bibinfo {author} {\bibfnamefont {B.}~\bibnamefont
  {Puchala}}, \bibinfo {author} {\bibfnamefont {M.~L.}\ \bibnamefont {Falk}}, \
  and\ \bibinfo {author} {\bibfnamefont {K.}~\bibnamefont {Garikipati}},\
  }\href {\doibase 10.1103/PhysRevB.77.174116} {\bibfield  {journal} {\bibinfo
  {journal} {Physical Review B - Condensed Matter and Materials Physics}\
  }\textbf {\bibinfo {volume} {77}},\ \bibinfo {pages} {1} (\bibinfo {year}
  {2008})},\ \Eprint {http://arxiv.org/abs/0802.1300} {arXiv:0802.1300}
  \BibitemShut {NoStop}%
\bibitem [{\citenamefont {Hinterberg}\ \emph {et~al.}(2013)\citenamefont
  {Hinterberg}, \citenamefont {Zacherle},\ and\ \citenamefont {{De
  Souza}}}]{Hinterberg2013}%
  \BibitemOpen
  \bibfield  {author} {\bibinfo {author} {\bibfnamefont {J.}~\bibnamefont
  {Hinterberg}}, \bibinfo {author} {\bibfnamefont {T.}~\bibnamefont
  {Zacherle}}, \ and\ \bibinfo {author} {\bibfnamefont {R.~a.}\ \bibnamefont
  {{De Souza}}},\ }\href {\doibase 10.1103/PhysRevLett.110.205901} {\bibfield
  {journal} {\bibinfo  {journal} {Physical Review Letters}\ }\textbf {\bibinfo
  {volume} {110}},\ \bibinfo {pages} {205901} (\bibinfo {year}
  {2013})}\BibitemShut {NoStop}%
\bibitem [{\citenamefont {Dorado}\ \emph {et~al.}(2011)\citenamefont {Dorado},
  \citenamefont {Garcia}, \citenamefont {Carlot}, \citenamefont {Davoisne},
  \citenamefont {Fraczkiewicz}, \citenamefont {Pasquet}, \citenamefont
  {Freyss}, \citenamefont {Valot}, \citenamefont {Baldinozzi}, \citenamefont
  {Sim{\'{e}}one},\ and\ \citenamefont {Bertolus}}]{Dorado2011}%
  \BibitemOpen
  \bibfield  {author} {\bibinfo {author} {\bibfnamefont {B.}~\bibnamefont
  {Dorado}}, \bibinfo {author} {\bibfnamefont {P.}~\bibnamefont {Garcia}},
  \bibinfo {author} {\bibfnamefont {G.}~\bibnamefont {Carlot}}, \bibinfo
  {author} {\bibfnamefont {C.}~\bibnamefont {Davoisne}}, \bibinfo {author}
  {\bibfnamefont {M.}~\bibnamefont {Fraczkiewicz}}, \bibinfo {author}
  {\bibfnamefont {B.}~\bibnamefont {Pasquet}}, \bibinfo {author} {\bibfnamefont
  {M.}~\bibnamefont {Freyss}}, \bibinfo {author} {\bibfnamefont
  {C.}~\bibnamefont {Valot}}, \bibinfo {author} {\bibfnamefont
  {G.}~\bibnamefont {Baldinozzi}}, \bibinfo {author} {\bibfnamefont
  {D.}~\bibnamefont {Sim{\'{e}}one}}, \ and\ \bibinfo {author} {\bibfnamefont
  {M.}~\bibnamefont {Bertolus}},\ }\href {\doibase 10.1103/PhysRevB.83.035126}
  {\bibfield  {journal} {\bibinfo  {journal} {Physical Review B}\ }\textbf
  {\bibinfo {volume} {83}},\ \bibinfo {pages} {035126} (\bibinfo {year}
  {2011})}\BibitemShut {NoStop}%
\bibitem [{\citenamefont {Andersson}\ \emph {et~al.}(2011)\citenamefont
  {Andersson}, \citenamefont {Uberuaga}, \citenamefont {Nerikar}, \citenamefont
  {Unal},\ and\ \citenamefont {Stanek}}]{Andersson2011}%
  \BibitemOpen
  \bibfield  {author} {\bibinfo {author} {\bibfnamefont {D.~a.}\ \bibnamefont
  {Andersson}}, \bibinfo {author} {\bibfnamefont {B.~P.}\ \bibnamefont
  {Uberuaga}}, \bibinfo {author} {\bibfnamefont {P.~V.}\ \bibnamefont
  {Nerikar}}, \bibinfo {author} {\bibfnamefont {C.}~\bibnamefont {Unal}}, \
  and\ \bibinfo {author} {\bibfnamefont {C.~R.}\ \bibnamefont {Stanek}},\
  }\href {\doibase 10.1103/PhysRevB.84.054105} {\bibfield  {journal} {\bibinfo
  {journal} {Physical Review B}\ }\textbf {\bibinfo {volume} {84}},\ \bibinfo
  {pages} {054105} (\bibinfo {year} {2011})}\BibitemShut {NoStop}%
\bibitem [{\citenamefont {Dorado}\ \emph {et~al.}(2012)\citenamefont {Dorado},
  \citenamefont {Andersson}, \citenamefont {Stanek}, \citenamefont {Bertolus},
  \citenamefont {Uberuaga}, \citenamefont {Martin}, \citenamefont {Freyss},\
  and\ \citenamefont {Garcia}}]{Dorado2012}%
  \BibitemOpen
  \bibfield  {author} {\bibinfo {author} {\bibfnamefont {B.}~\bibnamefont
  {Dorado}}, \bibinfo {author} {\bibfnamefont {D.}~\bibnamefont {Andersson}},
  \bibinfo {author} {\bibfnamefont {C.}~\bibnamefont {Stanek}}, \bibinfo
  {author} {\bibfnamefont {M.}~\bibnamefont {Bertolus}}, \bibinfo {author}
  {\bibfnamefont {B.}~\bibnamefont {Uberuaga}}, \bibinfo {author}
  {\bibfnamefont {G.}~\bibnamefont {Martin}}, \bibinfo {author} {\bibfnamefont
  {M.}~\bibnamefont {Freyss}}, \ and\ \bibinfo {author} {\bibfnamefont
  {P.}~\bibnamefont {Garcia}},\ }\href {\doibase 10.1103/PhysRevB.86.035110}
  {\bibfield  {journal} {\bibinfo  {journal} {Physical Review B}\ }\textbf
  {\bibinfo {volume} {86}},\ \bibinfo {pages} {035110} (\bibinfo {year}
  {2012})}\BibitemShut {NoStop}%
\bibitem [{\citenamefont {Nerikar}\ \emph {et~al.}(2009)\citenamefont
  {Nerikar}, \citenamefont {Liu}, \citenamefont {Uberuaga}, \citenamefont
  {Stanek}, \citenamefont {Phillpot},\ and\ \citenamefont
  {Sinnott}}]{Nerikar2009}%
  \BibitemOpen
  \bibfield  {author} {\bibinfo {author} {\bibfnamefont {P.~V.}\ \bibnamefont
  {Nerikar}}, \bibinfo {author} {\bibfnamefont {X.-Y.}\ \bibnamefont {Liu}},
  \bibinfo {author} {\bibfnamefont {B.~P.}\ \bibnamefont {Uberuaga}}, \bibinfo
  {author} {\bibfnamefont {C.~R.}\ \bibnamefont {Stanek}}, \bibinfo {author}
  {\bibfnamefont {S.~R.}\ \bibnamefont {Phillpot}}, \ and\ \bibinfo {author}
  {\bibfnamefont {S.~B.}\ \bibnamefont {Sinnott}},\ }\href {\doibase
  10.1088/0953-8984/21/43/435602} {\bibfield  {journal} {\bibinfo  {journal}
  {Journal of physics. Condensed matter : an Institute of Physics journal}\
  }\textbf {\bibinfo {volume} {21}},\ \bibinfo {pages} {435602} (\bibinfo
  {year} {2009})}\BibitemShut {NoStop}%
\bibitem [{\citenamefont {Kresse}\ and\ \citenamefont
  {Hafner}(1993)}]{Kresse1993}%
  \BibitemOpen
  \bibfield  {author} {\bibinfo {author} {\bibfnamefont {G.}~\bibnamefont
  {Kresse}}\ and\ \bibinfo {author} {\bibfnamefont {J.}~\bibnamefont
  {Hafner}},\ }\href {\doibase 10.1103/PhysRevB.47.558} {\bibfield  {journal}
  {\bibinfo  {journal} {Physical Review B}\ }\textbf {\bibinfo {volume} {47}},\
  \bibinfo {pages} {558} (\bibinfo {year} {1993})}\BibitemShut {NoStop}%
\bibitem [{\citenamefont {Kresse}\ and\ \citenamefont
  {Furthm{\"{u}}ller}(1996)}]{Kresse1996}%
  \BibitemOpen
  \bibfield  {author} {\bibinfo {author} {\bibfnamefont {G.}~\bibnamefont
  {Kresse}}\ and\ \bibinfo {author} {\bibfnamefont {J.}~\bibnamefont
  {Furthm{\"{u}}ller}},\ }\href {\doibase 10.1103/PhysRevB.54.11169} {\enquote
  {\bibinfo {title} {{Efficient iterative schemes for ab initio total-energy
  calculations using a plane-wave basis set}},}\ } (\bibinfo {year}
  {1996})\BibitemShut {NoStop}%
\bibitem [{\citenamefont {Bl{\"{o}}chl}(1994)}]{Blochl1994}%
  \BibitemOpen
  \bibfield  {author} {\bibinfo {author} {\bibfnamefont {P.~E.}\ \bibnamefont
  {Bl{\"{o}}chl}},\ }\href {\doibase 10.1103/PhysRevB.50.17953} {\bibfield
  {journal} {\bibinfo  {journal} {Physical Review B}\ }\textbf {\bibinfo
  {volume} {50}},\ \bibinfo {pages} {17953} (\bibinfo {year}
  {1994})}\BibitemShut {NoStop}%
\bibitem [{\citenamefont {Kresse}(1999)}]{Kresse1999}%
  \BibitemOpen
  \bibfield  {author} {\bibinfo {author} {\bibfnamefont {G.}~\bibnamefont
  {Kresse}},\ }\href {\doibase 10.1103/PhysRevB.59.1758} {\bibfield  {journal}
  {\bibinfo  {journal} {Physical Review B}\ }\textbf {\bibinfo {volume} {59}},\
  \bibinfo {pages} {1758} (\bibinfo {year} {1999})}\BibitemShut {NoStop}%
\bibitem [{\citenamefont {Perdew}\ and\ \citenamefont
  {Zunger}(1981)}]{Perdew1981}%
  \BibitemOpen
  \bibfield  {author} {\bibinfo {author} {\bibfnamefont {J.~P.}\ \bibnamefont
  {Perdew}}\ and\ \bibinfo {author} {\bibfnamefont {A.}~\bibnamefont
  {Zunger}},\ }\href {\doibase 10.1103/PhysRevB.23.5048} {\bibfield  {journal}
  {\bibinfo  {journal} {Physical Review B}\ }\textbf {\bibinfo {volume} {23}},\
  \bibinfo {pages} {5048} (\bibinfo {year} {1981})},\ \Eprint
  {http://arxiv.org/abs/0706.3359} {arXiv:0706.3359} \BibitemShut {NoStop}%
\bibitem [{\citenamefont {Liechtenstein}\ and\ \citenamefont
  {Zaanen}(1995)}]{Liechtenstein1995}%
  \BibitemOpen
  \bibfield  {author} {\bibinfo {author} {\bibfnamefont {A.~I.}\ \bibnamefont
  {Liechtenstein}}\ and\ \bibinfo {author} {\bibfnamefont {J.}~\bibnamefont
  {Zaanen}},\ }\href {\doibase 10.1103/PhysRevB.52.R5467} {\bibfield  {journal}
  {\bibinfo  {journal} {Physical Review B}\ }\textbf {\bibinfo {volume} {52}},\
  \bibinfo {pages} {R5467} (\bibinfo {year} {1995})}\BibitemShut {NoStop}%
\bibitem [{\citenamefont {Perdew}\ \emph {et~al.}(1996)\citenamefont {Perdew},
  \citenamefont {Burke},\ and\ \citenamefont {Ernzerhof}}]{Perdew1996}%
  \BibitemOpen
  \bibfield  {author} {\bibinfo {author} {\bibfnamefont {J.}~\bibnamefont
  {Perdew}}, \bibinfo {author} {\bibfnamefont {K.}~\bibnamefont {Burke}}, \
  and\ \bibinfo {author} {\bibfnamefont {M.}~\bibnamefont {Ernzerhof}},\ }\href
  {http://www.ncbi.nlm.nih.gov/pubmed/10062328} {\bibfield  {journal} {\bibinfo
   {journal} {Physical review letters}\ }\textbf {\bibinfo {volume} {77}},\
  \bibinfo {pages} {3865} (\bibinfo {year} {1996})}\BibitemShut {NoStop}%
\bibitem [{\citenamefont {Kumagai}\ and\ \citenamefont
  {Oba}(2014)}]{Kumagai2014}%
  \BibitemOpen
  \bibfield  {author} {\bibinfo {author} {\bibfnamefont {Y.}~\bibnamefont
  {Kumagai}}\ and\ \bibinfo {author} {\bibfnamefont {F.}~\bibnamefont {Oba}},\
  }\href {\doibase 10.1103/PhysRevB.89.195205} {\bibfield  {journal} {\bibinfo
  {journal} {Physical Review B}\ }\textbf {\bibinfo {volume} {89}},\ \bibinfo
  {pages} {195205} (\bibinfo {year} {2014})},\ \Eprint
  {http://arxiv.org/abs/1402.1226} {arXiv:1402.1226} \BibitemShut {NoStop}%
\bibitem [{\citenamefont {Freysoldt}\ \emph
  {et~al.}(2011{\natexlab{a}})\citenamefont {Freysoldt}, \citenamefont
  {Neugebauer},\ and\ \citenamefont {{Van de Walle}}}]{Freysoldt2011a}%
  \BibitemOpen
  \bibfield  {author} {\bibinfo {author} {\bibfnamefont {C.}~\bibnamefont
  {Freysoldt}}, \bibinfo {author} {\bibfnamefont {J.}~\bibnamefont
  {Neugebauer}}, \ and\ \bibinfo {author} {\bibfnamefont {C.~G.}\ \bibnamefont
  {{Van de Walle}}},\ }\href {\doibase 10.1002/pssb.201046289} {\bibfield
  {journal} {\bibinfo  {journal} {physica status solidi (b)}\ }\textbf
  {\bibinfo {volume} {248}},\ \bibinfo {pages} {1067} (\bibinfo {year}
  {2011}{\natexlab{a}})}\BibitemShut {NoStop}%
\bibitem [{\citenamefont {Baroni}\ and\ \citenamefont
  {Resta}(1986)}]{Baroni1986}%
  \BibitemOpen
  \bibfield  {author} {\bibinfo {author} {\bibfnamefont {S.}~\bibnamefont
  {Baroni}}\ and\ \bibinfo {author} {\bibfnamefont {R.}~\bibnamefont {Resta}},\
  }\href {\doibase 10.1103/PhysRevB.33.7017} {\bibfield  {journal} {\bibinfo
  {journal} {Physical Review B}\ }\textbf {\bibinfo {volume} {33}},\ \bibinfo
  {pages} {7017} (\bibinfo {year} {1986})}\BibitemShut {NoStop}%
\bibitem [{\citenamefont {Gajdo{\v{s}}}\ \emph {et~al.}(2006)\citenamefont
  {Gajdo{\v{s}}}, \citenamefont {Hummer}, \citenamefont {Kresse}, \citenamefont
  {Furthm{\"{u}}ller},\ and\ \citenamefont {Bechstedt}}]{Gajdos2006}%
  \BibitemOpen
  \bibfield  {author} {\bibinfo {author} {\bibfnamefont {M.}~\bibnamefont
  {Gajdo{\v{s}}}}, \bibinfo {author} {\bibfnamefont {K.}~\bibnamefont
  {Hummer}}, \bibinfo {author} {\bibfnamefont {G.}~\bibnamefont {Kresse}},
  \bibinfo {author} {\bibfnamefont {J.}~\bibnamefont {Furthm{\"{u}}ller}}, \
  and\ \bibinfo {author} {\bibfnamefont {F.}~\bibnamefont {Bechstedt}},\ }\href
  {\doibase 10.1103/PhysRevB.73.045112} {\bibfield  {journal} {\bibinfo
  {journal} {Physical Review B - Condensed Matter and Materials Physics}\
  }\textbf {\bibinfo {volume} {73}},\ \bibinfo {pages} {1} (\bibinfo {year}
  {2006})}\BibitemShut {NoStop}%
\bibitem [{\citenamefont {Sanati}\ \emph {et~al.}(2011)\citenamefont {Sanati},
  \citenamefont {Albers}, \citenamefont {Lookman},\ and\ \citenamefont
  {Saxena}}]{Sanati2011}%
  \BibitemOpen
  \bibfield  {author} {\bibinfo {author} {\bibfnamefont {M.}~\bibnamefont
  {Sanati}}, \bibinfo {author} {\bibfnamefont {R.}~\bibnamefont {Albers}},
  \bibinfo {author} {\bibfnamefont {T.}~\bibnamefont {Lookman}}, \ and\
  \bibinfo {author} {\bibfnamefont {a.}~\bibnamefont {Saxena}},\ }\href
  {\doibase 10.1103/PhysRevB.84.014116} {\bibfield  {journal} {\bibinfo
  {journal} {Physical Review B}\ }\textbf {\bibinfo {volume} {84}},\ \bibinfo
  {pages} {014116} (\bibinfo {year} {2011})}\BibitemShut {NoStop}%
\bibitem [{\citenamefont {Hampton}\ \emph {et~al.}(1987)\citenamefont
  {Hampton}, \citenamefont {Saunders}, \citenamefont {Harding},\ and\
  \citenamefont {Stoneham}}]{Hampton1987a}%
  \BibitemOpen
  \bibfield  {author} {\bibinfo {author} {\bibfnamefont {N.}~\bibnamefont
  {Hampton}}, \bibinfo {author} {\bibfnamefont {G.~A.}\ \bibnamefont
  {Saunders}}, \bibinfo {author} {\bibfnamefont {J.~H.}\ \bibnamefont
  {Harding}}, \ and\ \bibinfo {author} {\bibfnamefont {A.~M.}\ \bibnamefont
  {Stoneham}},\ }\href {\doibase 10.1016/0022-3115(87)90493-4} {\bibfield
  {journal} {\bibinfo  {journal} {Journal of Nuclear Materials}\ }\textbf
  {\bibinfo {volume} {149}},\ \bibinfo {pages} {18} (\bibinfo {year}
  {1987})}\BibitemShut {NoStop}%
\bibitem [{\citenamefont {Lee}\ and\ \citenamefont {Rudd}(2007)}]{Lee2007}%
  \BibitemOpen
  \bibfield  {author} {\bibinfo {author} {\bibfnamefont {B.}~\bibnamefont
  {Lee}}\ and\ \bibinfo {author} {\bibfnamefont {R.~E.}\ \bibnamefont {Rudd}},\
  }\href {\doibase 10.1103/PhysRevB.75.195328} {\bibfield  {journal} {\bibinfo
  {journal} {Physical Review B}\ }\textbf {\bibinfo {volume} {75}},\ \bibinfo
  {pages} {195328} (\bibinfo {year} {2007})},\ \Eprint
  {http://arxiv.org/abs/0702531} {arXiv:0702531 [cond-mat]} \BibitemShut
  {NoStop}%
\bibitem [{\citenamefont {McSkimin}(1953)}]{McSkimin1953}%
  \BibitemOpen
  \bibfield  {author} {\bibinfo {author} {\bibfnamefont {H.~J.}\ \bibnamefont
  {McSkimin}},\ }\href {\doibase 10.1063/1.1721449} {\bibfield  {journal}
  {\bibinfo  {journal} {Journal of Applied Physics}\ }\textbf {\bibinfo
  {volume} {24}},\ \bibinfo {pages} {988} (\bibinfo {year} {1953})}\BibitemShut
  {NoStop}%
\bibitem [{\citenamefont {Madelung}(1991)}]{Madelung1991}%
  \BibitemOpen
  \bibfield  {author} {\bibinfo {author} {\bibfnamefont {O.}~\bibnamefont
  {Madelung}},\ }\href {file://catalog.hathitrust.org/Record/002480151
  http://hdl.handle.net/2027/mdp.39015021980738
  http://hdl.handle.net/2027/mdp.39015021506764} {\enquote {\bibinfo {title}
  {{Semiconductors: group IV elements and III-V compounds}},}\ } (\bibinfo
  {year} {1991})\BibitemShut {NoStop}%
\bibitem [{\citenamefont {Komsa}\ \emph {et~al.}(2012)\citenamefont {Komsa},
  \citenamefont {Rantala},\ and\ \citenamefont {Pasquarello}}]{Komsa2012}%
  \BibitemOpen
  \bibfield  {author} {\bibinfo {author} {\bibfnamefont {H.-P.}\ \bibnamefont
  {Komsa}}, \bibinfo {author} {\bibfnamefont {T.~T.}\ \bibnamefont {Rantala}},
  \ and\ \bibinfo {author} {\bibfnamefont {A.}~\bibnamefont {Pasquarello}},\
  }\href {\doibase 10.1103/PhysRevB.86.045112} {\bibfield  {journal} {\bibinfo
  {journal} {Physical Review B}\ }\textbf {\bibinfo {volume} {86}},\ \bibinfo
  {pages} {045112} (\bibinfo {year} {2012})}\BibitemShut {NoStop}%
\bibitem [{\citenamefont {Cooper}(1962)}]{Cooper1962}%
  \BibitemOpen
  \bibfield  {author} {\bibinfo {author} {\bibfnamefont {A.~S.}\ \bibnamefont
  {Cooper}},\ }\href {\doibase 10.1107/S0365110X62001474} {\bibfield  {journal}
  {\bibinfo  {journal} {Acta Crystallographica}\ }\textbf {\bibinfo {volume}
  {15}},\ \bibinfo {pages} {578} (\bibinfo {year} {1962})}\BibitemShut
  {NoStop}%
\bibitem [{\citenamefont {Garland}\ and\ \citenamefont
  {Park}(1962)}]{Garland1962}%
  \BibitemOpen
  \bibfield  {author} {\bibinfo {author} {\bibfnamefont {C.~W.}\ \bibnamefont
  {Garland}}\ and\ \bibinfo {author} {\bibfnamefont {K.~C.}\ \bibnamefont
  {Park}},\ }\href {\doibase 10.1063/1.1702519} {\bibfield  {journal} {\bibinfo
   {journal} {Journal of Applied Physics}\ }\textbf {\bibinfo {volume} {33}},\
  \bibinfo {pages} {759} (\bibinfo {year} {1962})}\BibitemShut {NoStop}%
\bibitem [{\citenamefont {Dorado}\ and\ \citenamefont
  {Garcia}(2013)}]{Dorado2013}%
  \BibitemOpen
  \bibfield  {author} {\bibinfo {author} {\bibfnamefont {B.}~\bibnamefont
  {Dorado}}\ and\ \bibinfo {author} {\bibfnamefont {P.}~\bibnamefont
  {Garcia}},\ }\href {\doibase 10.1103/PhysRevB.87.195139} {\bibfield
  {journal} {\bibinfo  {journal} {Physical Review B}\ }\textbf {\bibinfo
  {volume} {87}},\ \bibinfo {pages} {195139} (\bibinfo {year}
  {2013})}\BibitemShut {NoStop}%
\bibitem [{\citenamefont {Clark}\ \emph {et~al.}(2005)\citenamefont {Clark},
  \citenamefont {Segall}, \citenamefont {Pickard}, \citenamefont {Hasnip},
  \citenamefont {Probert}, \citenamefont {Refson},\ and\ \citenamefont
  {Payne}}]{Clark2005}%
  \BibitemOpen
  \bibfield  {author} {\bibinfo {author} {\bibfnamefont {S.~J.}\ \bibnamefont
  {Clark}}, \bibinfo {author} {\bibfnamefont {M.~D.}\ \bibnamefont {Segall}},
  \bibinfo {author} {\bibfnamefont {C.~J.}\ \bibnamefont {Pickard}}, \bibinfo
  {author} {\bibfnamefont {P.~J.}\ \bibnamefont {Hasnip}}, \bibinfo {author}
  {\bibfnamefont {M.~I.~J.}\ \bibnamefont {Probert}}, \bibinfo {author}
  {\bibfnamefont {K.}~\bibnamefont {Refson}}, \ and\ \bibinfo {author}
  {\bibfnamefont {M.~C.}\ \bibnamefont {Payne}},\ }\href {\doibase
  10.1524/zkri.220.5.567.65075} {\bibfield  {journal} {\bibinfo  {journal}
  {Zeitschrift f{\"{u}}r Kristallographie - Crystalline Materials}\ }\textbf
  {\bibinfo {volume} {220}} (\bibinfo {year} {2005}),\
  10.1524/zkri.220.5.567.65075}\BibitemShut {NoStop}%
\bibitem [{\citenamefont {Freysoldt}\ \emph {et~al.}(2014)\citenamefont
  {Freysoldt}, \citenamefont {Grabowski}, \citenamefont {Hickel}, \citenamefont
  {Neugebauer}, \citenamefont {Kresse}, \citenamefont {Janotti},\ and\
  \citenamefont {{Van de Walle}}}]{Freysoldt2014}%
  \BibitemOpen
  \bibfield  {author} {\bibinfo {author} {\bibfnamefont {C.}~\bibnamefont
  {Freysoldt}}, \bibinfo {author} {\bibfnamefont {B.}~\bibnamefont
  {Grabowski}}, \bibinfo {author} {\bibfnamefont {T.}~\bibnamefont {Hickel}},
  \bibinfo {author} {\bibfnamefont {J.}~\bibnamefont {Neugebauer}}, \bibinfo
  {author} {\bibfnamefont {G.}~\bibnamefont {Kresse}}, \bibinfo {author}
  {\bibfnamefont {A.}~\bibnamefont {Janotti}}, \ and\ \bibinfo {author}
  {\bibfnamefont {C.~G.}\ \bibnamefont {{Van de Walle}}},\ }\href {\doibase
  10.1103/RevModPhys.86.253} {\bibfield  {journal} {\bibinfo  {journal}
  {Reviews of Modern Physics}\ }\textbf {\bibinfo {volume} {86}},\ \bibinfo
  {pages} {253} (\bibinfo {year} {2014})}\BibitemShut {NoStop}%
\bibitem [{\citenamefont {Freysoldt}\ \emph {et~al.}(2009)\citenamefont
  {Freysoldt}, \citenamefont {Neugebauer},\ and\ \citenamefont
  {de~Walle}}]{Freysoldt2009}%
  \BibitemOpen
  \bibfield  {author} {\bibinfo {author} {\bibfnamefont {C.}~\bibnamefont
  {Freysoldt}}, \bibinfo {author} {\bibfnamefont {J.}~\bibnamefont
  {Neugebauer}}, \ and\ \bibinfo {author} {\bibfnamefont {C.~V.}\ \bibnamefont
  {de~Walle}},\ }\href {\doibase 10.1103/PhysRevLett.102.016402} {\bibfield
  {journal} {\bibinfo  {journal} {Physical review letters}\ }\textbf {\bibinfo
  {volume} {016402}},\ \bibinfo {pages} {1} (\bibinfo {year}
  {2009})}\BibitemShut {NoStop}%
\bibitem [{\citenamefont {Garikipati}\ \emph {et~al.}(2006)\citenamefont
  {Garikipati}, \citenamefont {Falk}, \citenamefont {Bouville}, \citenamefont
  {Puchala},\ and\ \citenamefont {Narayanan}}]{Garikipati2006}%
  \BibitemOpen
  \bibfield  {author} {\bibinfo {author} {\bibfnamefont {K.}~\bibnamefont
  {Garikipati}}, \bibinfo {author} {\bibfnamefont {M.}~\bibnamefont {Falk}},
  \bibinfo {author} {\bibfnamefont {M.}~\bibnamefont {Bouville}}, \bibinfo
  {author} {\bibfnamefont {B.}~\bibnamefont {Puchala}}, \ and\ \bibinfo
  {author} {\bibfnamefont {H.}~\bibnamefont {Narayanan}},\ }\href {\doibase
  10.1016/j.jmps.2006.02.007} {\bibfield  {journal} {\bibinfo  {journal}
  {Journal of the Mechanics and Physics of Solids}\ }\textbf {\bibinfo {volume}
  {54}},\ \bibinfo {pages} {1929} (\bibinfo {year} {2006})},\ \Eprint
  {http://arxiv.org/abs/0508169} {arXiv:0508169 [cond-mat]} \BibitemShut
  {NoStop}%
\bibitem [{\citenamefont {Desgranges}\ \emph {et~al.}(2009)\citenamefont
  {Desgranges}, \citenamefont {Baldinozzi}, \citenamefont {Rousseau},
  \citenamefont {Nièpce},\ and\ \citenamefont {Calvarin}}]{Desgranges2009}%
  \BibitemOpen
  \bibfield  {author} {\bibinfo {author} {\bibfnamefont {L.}~\bibnamefont
  {Desgranges}}, \bibinfo {author} {\bibfnamefont {G.}~\bibnamefont
  {Baldinozzi}}, \bibinfo {author} {\bibfnamefont {G.}~\bibnamefont
  {Rousseau}}, \bibinfo {author} {\bibfnamefont {J.-C.}\ \bibnamefont
  {Nièpce}}, \ and\ \bibinfo {author} {\bibfnamefont {G.}~\bibnamefont
  {Calvarin}},\ }\href {\doibase 10.1021/ic9000889} {\bibfield  {journal}
  {\bibinfo  {journal} {Inorganic Chemistry}\ }\textbf {\bibinfo {volume}
  {48}},\ \bibinfo {pages} {7585} (\bibinfo {year} {2009})}\BibitemShut
  {NoStop}%
\bibitem [{\citenamefont {Heine}(1968)}]{Heine1968}%
  \BibitemOpen
  \bibfield  {author} {\bibinfo {author} {\bibfnamefont {V.}~\bibnamefont
  {Heine}},\ }\href@noop {} {\bibfield  {journal} {\bibinfo  {journal} {Journal
  of Physics C}\ }\textbf {\bibinfo {volume} {1}},\ \bibinfo {pages} {222}
  (\bibinfo {year} {1968})}\BibitemShut {NoStop}%
\bibitem [{\citenamefont {Schiferl}\ and\ \citenamefont
  {Barrett}(1969)}]{Schiferl1969}%
  \BibitemOpen
  \bibfield  {author} {\bibinfo {author} {\bibfnamefont {D.}~\bibnamefont
  {Schiferl}}\ and\ \bibinfo {author} {\bibfnamefont {C.~S.}\ \bibnamefont
  {Barrett}},\ }\href {\doibase 10.1107/S0021889869006443} {\bibfield
  {journal} {\bibinfo  {journal} {Journal of Applied Crystallography}\ }\textbf
  {\bibinfo {volume} {2}},\ \bibinfo {pages} {30} (\bibinfo {year}
  {1969})}\BibitemShut {NoStop}%
\bibitem [{\citenamefont {Barrett}\ \emph {et~al.}(1963)\citenamefont
  {Barrett}, \citenamefont {Mueller},\ and\ \citenamefont
  {Hitterman}}]{Barrett1963}%
  \BibitemOpen
  \bibfield  {author} {\bibinfo {author} {\bibfnamefont {C.~S.}\ \bibnamefont
  {Barrett}}, \bibinfo {author} {\bibfnamefont {M.~H.}\ \bibnamefont
  {Mueller}}, \ and\ \bibinfo {author} {\bibfnamefont {R.~L.}\ \bibnamefont
  {Hitterman}},\ }\href {\doibase 10.1103/PhysRev.129.625} {\bibfield
  {journal} {\bibinfo  {journal} {Physical Review}\ }\textbf {\bibinfo {volume}
  {129}},\ \bibinfo {pages} {625} (\bibinfo {year} {1963})}\BibitemShut
  {NoStop}%
\bibitem [{\citenamefont {Allen}\ and\ \citenamefont
  {Holmes}(1995)}]{Allen1995}%
  \BibitemOpen
  \bibfield  {author} {\bibinfo {author} {\bibfnamefont {G.}~\bibnamefont
  {Allen}}\ and\ \bibinfo {author} {\bibfnamefont {N.}~\bibnamefont {Holmes}},\
  }\href {\doibase 10.1016/0022-3115(95)00025-9} {\bibfield  {journal}
  {\bibinfo  {journal} {Journal of Nuclear Materials}\ }\textbf {\bibinfo
  {volume} {223}},\ \bibinfo {pages} {231} (\bibinfo {year}
  {1995})}\BibitemShut {NoStop}%
\bibitem [{\citenamefont {Goyal}\ \emph {et~al.}(2013)\citenamefont {Goyal},
  \citenamefont {Rudzik}, \citenamefont {Deng}, \citenamefont {Hong},
  \citenamefont {Chernatynskiy}, \citenamefont {Sinnott},\ and\ \citenamefont
  {Phillpot}}]{Goyal2013}%
  \BibitemOpen
  \bibfield  {author} {\bibinfo {author} {\bibfnamefont {A.}~\bibnamefont
  {Goyal}}, \bibinfo {author} {\bibfnamefont {T.}~\bibnamefont {Rudzik}},
  \bibinfo {author} {\bibfnamefont {B.}~\bibnamefont {Deng}}, \bibinfo {author}
  {\bibfnamefont {M.}~\bibnamefont {Hong}}, \bibinfo {author} {\bibfnamefont
  {A.}~\bibnamefont {Chernatynskiy}}, \bibinfo {author} {\bibfnamefont {S.~B.}\
  \bibnamefont {Sinnott}}, \ and\ \bibinfo {author} {\bibfnamefont {S.~R.}\
  \bibnamefont {Phillpot}},\ }\href {\doibase 10.1016/j.jnucmat.2013.05.031}
  {\bibfield  {journal} {\bibinfo  {journal} {Journal of Nuclear Materials}\
  }\textbf {\bibinfo {volume} {441}},\ \bibinfo {pages} {96} (\bibinfo {year}
  {2013})}\BibitemShut {NoStop}%
\bibitem [{\citenamefont {Makov}\ and\ \citenamefont
  {Payne}(1995)}]{Makov1995}%
  \BibitemOpen
  \bibfield  {author} {\bibinfo {author} {\bibfnamefont {G.}~\bibnamefont
  {Makov}}\ and\ \bibinfo {author} {\bibfnamefont {M.}~\bibnamefont {Payne}},\
  }\href {\doibase 10.1103/PhysRevB.51.4014} {\bibfield  {journal} {\bibinfo
  {journal} {Physical Review B}\ }\textbf {\bibinfo {volume} {51}},\ \bibinfo
  {pages} {4014} (\bibinfo {year} {1995})}\BibitemShut {NoStop}%
\bibitem [{\citenamefont {Lany}\ and\ \citenamefont {Zunger}(2008)}]{Lany2008}%
  \BibitemOpen
  \bibfield  {author} {\bibinfo {author} {\bibfnamefont {S.}~\bibnamefont
  {Lany}}\ and\ \bibinfo {author} {\bibfnamefont {A.}~\bibnamefont {Zunger}},\
  }\href {\doibase 10.1103/PhysRevB.78.235104} {\bibfield  {journal} {\bibinfo
  {journal} {Physical Review B}\ }\textbf {\bibinfo {volume} {78}},\ \bibinfo
  {pages} {235104} (\bibinfo {year} {2008})}\BibitemShut {NoStop}%
\bibitem [{\citenamefont {Freysoldt}\ \emph
  {et~al.}(2011{\natexlab{b}})\citenamefont {Freysoldt}, \citenamefont
  {Neugebauer},\ and\ \citenamefont {{Van de Walle}}}]{Freysoldt2011}%
  \BibitemOpen
  \bibfield  {author} {\bibinfo {author} {\bibfnamefont {C.}~\bibnamefont
  {Freysoldt}}, \bibinfo {author} {\bibfnamefont {J.}~\bibnamefont
  {Neugebauer}}, \ and\ \bibinfo {author} {\bibfnamefont {C.~G.}\ \bibnamefont
  {{Van de Walle}}},\ }\href {\doibase 10.1002/pssb.201046289} {\bibfield
  {journal} {\bibinfo  {journal} {Physica Status Solidi (B)}\ }\textbf
  {\bibinfo {volume} {248}},\ \bibinfo {pages} {1067} (\bibinfo {year}
  {2011}{\natexlab{b}})}\BibitemShut {NoStop}%
\bibitem [{\citenamefont {Leslie}\ and\ \citenamefont
  {Gillan}(1985)}]{Leslie1985}%
  \BibitemOpen
  \bibfield  {author} {\bibinfo {author} {\bibfnamefont {M.}~\bibnamefont
  {Leslie}}\ and\ \bibinfo {author} {\bibfnamefont {M.~J.}\ \bibnamefont
  {Gillan}},\ }\href {http://iopscience.iop.org/0022-3719/18/5/005} {\bibfield
  {journal} {\bibinfo  {journal} {Journal of Physics C: Solid State Physics}\
  }\textbf {\bibinfo {volume} {18}},\ \bibinfo {pages} {973} (\bibinfo {year}
  {1985})}\BibitemShut {NoStop}%
\bibitem [{\citenamefont {Murphy}\ and\ \citenamefont
  {Hine}(2013)}]{Murphy2013}%
  \BibitemOpen
  \bibfield  {author} {\bibinfo {author} {\bibfnamefont {S.~T.}\ \bibnamefont
  {Murphy}}\ and\ \bibinfo {author} {\bibfnamefont {N.~D.~M.}\ \bibnamefont
  {Hine}},\ }\href {\doibase 10.1103/PhysRevB.87.094111} {\bibfield  {journal}
  {\bibinfo  {journal} {Physical Review B - Condensed Matter and Materials
  Physics}\ }\textbf {\bibinfo {volume} {87}},\ \bibinfo {pages} {1} (\bibinfo
  {year} {2013})},\ \Eprint {http://arxiv.org/abs/1303.5377} {arXiv:1303.5377}
  \BibitemShut {NoStop}%
\bibitem [{\citenamefont {Taylor}\ and\ \citenamefont
  {Bruneval}(2011)}]{Taylor2011}%
  \BibitemOpen
  \bibfield  {author} {\bibinfo {author} {\bibfnamefont {S.~E.}\ \bibnamefont
  {Taylor}}\ and\ \bibinfo {author} {\bibfnamefont {F.}~\bibnamefont
  {Bruneval}},\ }\href {\doibase 10.1103/PhysRevB.84.075155} {\bibfield
  {journal} {\bibinfo  {journal} {Physical Review B}\ }\textbf {\bibinfo
  {volume} {84}},\ \bibinfo {pages} {075155} (\bibinfo {year}
  {2011})}\BibitemShut {NoStop}%
\bibitem [{\citenamefont {Windl}\ and\ \citenamefont
  {Daw}(2002{\natexlab{b}})}]{Windl2002}%
  \BibitemOpen
  \bibfield  {author} {\bibinfo {author} {\bibfnamefont {W.}~\bibnamefont
  {Windl}}\ and\ \bibinfo {author} {\bibfnamefont {M.}~\bibnamefont {Daw}},\
  }in\ \href@noop {} {\emph {\bibinfo {booktitle} {International Conference on
  Modeling and SImulatuion of Microsystems - MSM 2002}}},\ \bibinfo {editor}
  {edited by\ \bibinfo {editor} {\bibfnamefont {M.}~\bibnamefont {Laudon}}\
  and\ \bibinfo {editor} {\bibfnamefont {B.}~\bibnamefont {Romanowicz}}}\
  (\bibinfo {year} {2002})\ pp.\ \bibinfo {pages} {494--497}\BibitemShut
  {NoStop}%
\bibitem [{\citenamefont {Bader}(1990)}]{Bader1990}%
  \BibitemOpen
  \bibfield  {author} {\bibinfo {author} {\bibfnamefont {R.~F.~W.}\
  \bibnamefont {Bader}},\ }\href
  {https://books.google.com/books?id=up1pQgAACAAJ} {\emph {\bibinfo {title}
  {{Atoms in Molecules: A Quantum Theory}}}},\ International series of
  monographs on chemistry\ (\bibinfo  {publisher} {Clarendon Press},\ \bibinfo
  {year} {1990})\BibitemShut {NoStop}%
\bibitem [{\citenamefont {Henkelman}\ \emph {et~al.}(2006)\citenamefont
  {Henkelman}, \citenamefont {Arnaldsson},\ and\ \citenamefont
  {J{\'{o}}nsson}}]{Henkelman2006}%
  \BibitemOpen
  \bibfield  {author} {\bibinfo {author} {\bibfnamefont {G.}~\bibnamefont
  {Henkelman}}, \bibinfo {author} {\bibfnamefont {A.}~\bibnamefont
  {Arnaldsson}}, \ and\ \bibinfo {author} {\bibfnamefont {H.}~\bibnamefont
  {J{\'{o}}nsson}},\ }\href {\doibase 10.1016/j.commatsci.2005.04.010}
  {\bibfield  {journal} {\bibinfo  {journal} {Computational Materials Science}\
  }\textbf {\bibinfo {volume} {36}},\ \bibinfo {pages} {354} (\bibinfo {year}
  {2006})}\BibitemShut {NoStop}%
\bibitem [{\citenamefont {Dorado}\ \emph {et~al.}(2010)\citenamefont {Dorado},
  \citenamefont {Jomard}, \citenamefont {Freyss},\ and\ \citenamefont
  {Bertolus}}]{Dorado2010}%
  \BibitemOpen
  \bibfield  {author} {\bibinfo {author} {\bibfnamefont {B.}~\bibnamefont
  {Dorado}}, \bibinfo {author} {\bibfnamefont {G.}~\bibnamefont {Jomard}},
  \bibinfo {author} {\bibfnamefont {M.}~\bibnamefont {Freyss}}, \ and\ \bibinfo
  {author} {\bibfnamefont {M.}~\bibnamefont {Bertolus}},\ }\href {\doibase
  10.1103/PhysRevB.82.035114} {\bibfield  {journal} {\bibinfo  {journal}
  {Physical Review B}\ }\textbf {\bibinfo {volume} {82}},\ \bibinfo {pages}
  {035114} (\bibinfo {year} {2010})}\BibitemShut {NoStop}%
\bibitem [{\citenamefont {Stoneham}\ \emph {et~al.}(2007)\citenamefont
  {Stoneham}, \citenamefont {Gavartin}, \citenamefont {Shluger}, \citenamefont
  {Kimmel}, \citenamefont {Ramo}, \citenamefont {R{\o}nnow}, \citenamefont
  {Aeppli},\ and\ \citenamefont {Renner}}]{Stoneham2007}%
  \BibitemOpen
  \bibfield  {author} {\bibinfo {author} {\bibfnamefont {A.~M.}\ \bibnamefont
  {Stoneham}}, \bibinfo {author} {\bibfnamefont {J.}~\bibnamefont {Gavartin}},
  \bibinfo {author} {\bibfnamefont {A.~L.}\ \bibnamefont {Shluger}}, \bibinfo
  {author} {\bibfnamefont {A.~V.}\ \bibnamefont {Kimmel}}, \bibinfo {author}
  {\bibfnamefont {D.~M.}\ \bibnamefont {Ramo}}, \bibinfo {author}
  {\bibfnamefont {H.~M.}\ \bibnamefont {R{\o}nnow}}, \bibinfo {author}
  {\bibfnamefont {G.}~\bibnamefont {Aeppli}}, \ and\ \bibinfo {author}
  {\bibfnamefont {C.}~\bibnamefont {Renner}},\ }\href {\doibase
  10.1088/0953-8984/19/25/255208} {\bibfield  {journal} {\bibinfo  {journal}
  {Journal of Physics: Condensed Matter}\ }\textbf {\bibinfo {volume} {19}},\
  \bibinfo {pages} {255208} (\bibinfo {year} {2007})}\BibitemShut {NoStop}%
\bibitem [{\citenamefont {Kr{\"{o}}ger}\ and\ \citenamefont
  {Vink}(1956)}]{Kroger1956}%
  \BibitemOpen
  \bibfield  {author} {\bibinfo {author} {\bibfnamefont {F.~A.}\ \bibnamefont
  {Kr{\"{o}}ger}}\ and\ \bibinfo {author} {\bibfnamefont {H.~J.}\ \bibnamefont
  {Vink}},\ }\href {\doibase 10.1016/S0081-1947(08)60135-6} {\bibfield
  {journal} {\bibinfo  {journal} {Solid State Physics - Advances in Research
  and Applications}\ }\textbf {\bibinfo {volume} {3}},\ \bibinfo {pages} {307}
  (\bibinfo {year} {1956})}\BibitemShut {NoStop}%
\bibitem [{\citenamefont {Crocombette}\ \emph {et~al.}(2001)\citenamefont
  {Crocombette}, \citenamefont {Jollet}, \citenamefont {Nga},\ and\
  \citenamefont {Petit}}]{Crocombette2001}%
  \BibitemOpen
  \bibfield  {author} {\bibinfo {author} {\bibfnamefont {J.}~\bibnamefont
  {Crocombette}}, \bibinfo {author} {\bibfnamefont {F.}~\bibnamefont {Jollet}},
  \bibinfo {author} {\bibfnamefont {L.}~\bibnamefont {Nga}}, \ and\ \bibinfo
  {author} {\bibfnamefont {T.}~\bibnamefont {Petit}},\ }\href {\doibase
  10.1103/PhysRevB.64.104107} {\bibfield  {journal} {\bibinfo  {journal}
  {Physical Review B}\ }\textbf {\bibinfo {volume} {64}},\ \bibinfo {pages}
  {104107} (\bibinfo {year} {2001})}\BibitemShut {NoStop}%
\end{thebibliography}%

\end{document}